\documentclass{article}
\usepackage[table]{xcolor}

\usepackage{microtype}
\usepackage{graphicx}
\usepackage{subfigure}
\usepackage{booktabs}
\usepackage{array}
\usepackage{hyperref}
\usepackage{xurl}
\usepackage{placeins}
\usepackage{float}
\usepackage{amssymb}
\usepackage{amsmath}
\usepackage{tabularx}
\definecolor{BenchNavy}{HTML}{660000}
\definecolor{BenchLight}{HTML}{FFF2F2}
\newcommand{\benchfinding}[2]{%
  \par\smallskip\noindent\begingroup
  \setlength{\fboxsep}{3pt}\setlength{\fboxrule}{1pt}%
  \vbox{\offinterlineskip
    \hbox{\colorbox{BenchNavy}{\parbox{\dimexpr\columnwidth-2\fboxsep\relax}{\color{white}\small\bfseries\raggedright\hyphenpenalty=10000 Insight #1}}}%
    \hbox{\fcolorbox{BenchNavy}{BenchLight}{\parbox{\dimexpr\columnwidth-2\fboxsep-2\fboxrule\relax}{\small #2}}}%
  }\endgroup\par\smallskip
}

\usepackage[accepted]{mlsys2025}
\makeatletter
\renewcommand{\Notice@String}{}
\makeatother
\hypersetup{pdfsubject={Preprint},
  pdftitle={MAServeBench: Benchmarking Multi-Agent LLM Serving Systems with Hierarchical Collaboration Workloads}}

\newenvironment{benchequation}{%
  \begingroup\small
  \setlength{\abovedisplayskip}{4pt}%
  \setlength{\belowdisplayskip}{4pt}%
  \setlength{\abovedisplayshortskip}{3pt}%
  \setlength{\belowdisplayshortskip}{3pt}%
  \begin{equation}%
}{\end{equation}\endgroup}

\mlsystitlerunning{MAServeBench: Benchmarking Multi-Agent LLM Serving Systems with Hierarchical Collaboration Workloads}

\begin{document}

\twocolumn[
\mlsystitle{MAServeBench: Benchmarking Multi-Agent LLM Serving Systems with Hierarchical Collaboration Workloads}

% Author order and institution assignments supplied by the authors.
\begin{mlsysauthorlist}
\mlsysauthor{Shengxuan Qiu}{pku}
\mlsysauthor{Haochen Huang}{pku}
\mlsysauthor{Shuzhang Zhong}{pku}
\mlsysauthor{Zizhuo Fu}{pku}
\mlsysauthor{Zebin Yang}{pku}
% \mlsysauthor{Jianming Tong}{gatech}
\mlsysauthor{Meng Li}{pku}
\end{mlsysauthorlist}

\mlsysaffiliation{pku}{Peking University}
% \mlsysaffiliation{gatech}{Georgia Institute of Technology}
\mlsyscorrespondingauthor{Meng Li}{meng.li@pku.edu.cn}
\mlsyskeywords{Multi-Agent Systems, LLM Serving, Benchmarking, Execution Traces}

\vskip 0.3in

\begin{abstract}
Large language model (LLM)-based multi-agent systems (MAS) coordinate specialized agents to solve complex tasks, turning each user request into interdependent model and tool operations. Unlike independent requests, these operations become ready at different times and consume evolving context as agents coordinate and exchange results. This creates changing patterns of concurrency, computation, and memory demand that can expose different bottlenecks in LLM serving hardware. Yet benchmarking these effects requires capturing the collaboration rules that generate serving workloads, beyond the requests observed during execution. We present \textbf{MAServeBench}, a benchmark that treats multi-agent collaboration as a configurable dimension of serving workloads for LLM serving and hardware evaluation. Its composable stages capture how agents coordinate and exchange context in documented MAS workflows, enabling native executions to generate different workloads even when task inputs and tools remain unchanged. Hierarchical Semantic Traces preserve realized operations and dependencies for reproducible fixed-work replay and as inputs to architecture simulators. Task-linked measurements connect these workloads to serving execution and GPU/NPU hardware behavior, revealing why serving efficiency does not always translate into timely task completion.
The benchmark code is available at \mbox{\nolinkurl{https://github.com/ShengxuanQiu/MAServeBench}}.

\end{abstract}
]

% This must follow the closing bracket of \twocolumn[...].
\printAffiliationsAndNotice{}

\section{Introduction}
\label{sec:introduction}

Large language model (LLM) agents are increasingly used to tackle tasks that require multi-step reasoning, tool use, and interaction with external environments~\cite{react,webarena}. As these tasks grow more complex, multi-agent systems (MAS) offer a way to divide work among agents with complementary roles. Instead of asking one agent to plan, gather information, act, and verify every step, MAS can organize these activities through specialization, parallel exploration, and iterative feedback. Systems such as AutoGen, MetaGPT, ChatDev, and Magentic-One demonstrate how agents can coordinate their contributions toward a shared task~\cite{autogen,metagpt,chatdev,magenticone}. 

\begin{figure}[t]
  \centering
  \includegraphics[width=\columnwidth]{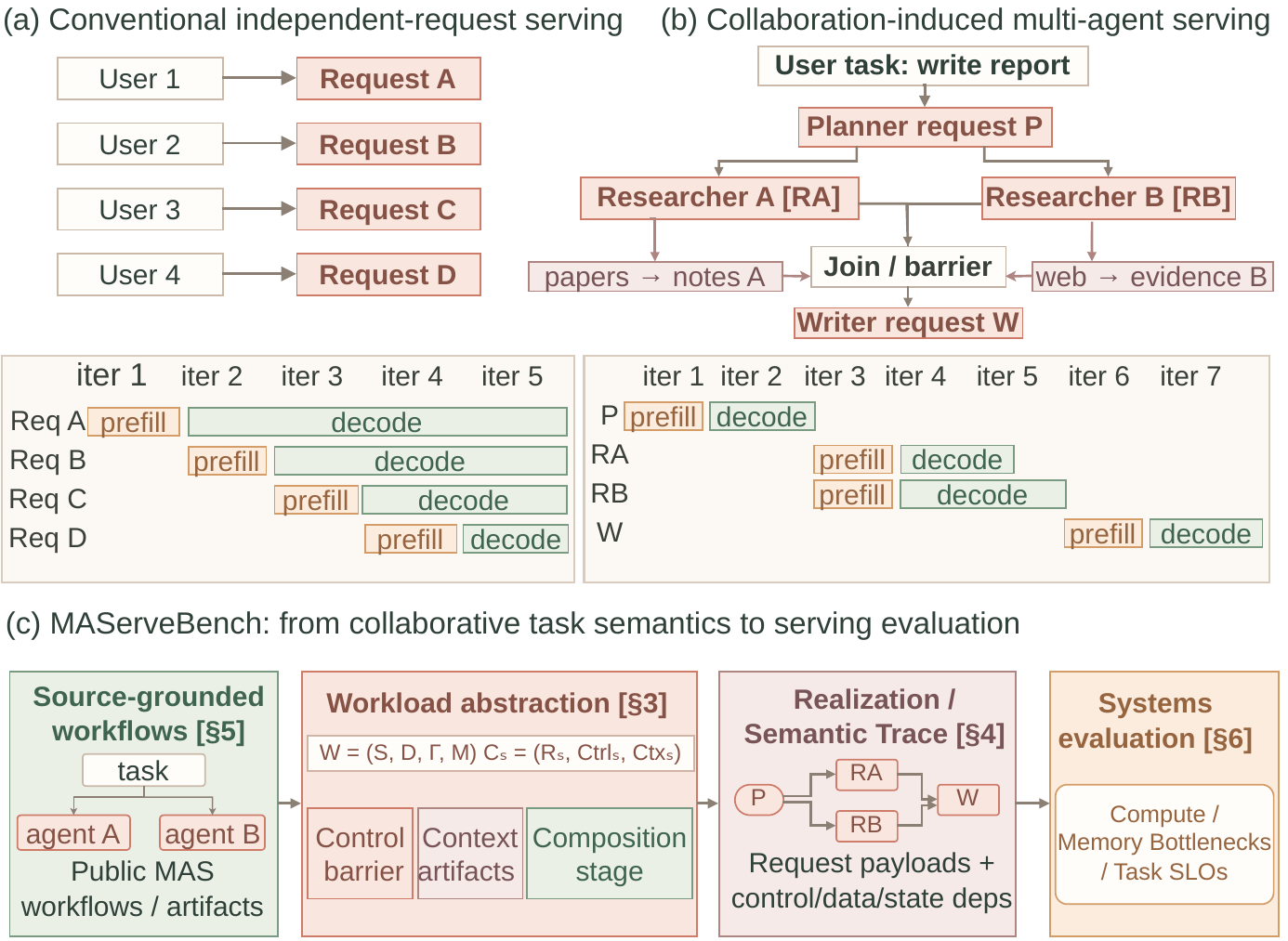}
  \vspace{-20pt}
  \caption{\textbf{Collaboration reshapes serving demand.} (a) Independent request arrivals. (b) A multi-agent task induces dependent requests with artifact exchange and synchronization. Both can use continuous batching. (c) MAServeBench enables controlled, reproducible evaluation of collaboration-induced serving demand and hardware resource behavior.}
  \label{fig:intro}
\end{figure}

However, the collaboration that makes MAS useful also reshapes LLM serving demand; its diverse, framework-specific patterns hinder reusable workload construction. Conventional serving treats model requests as independent arrivals, while a single-agent workflow typically follows one evolving interaction. In MAS, requests may depend on other agents' progress and outputs: parallel branches may release several calls together, synchronization may delay downstream calls, and refinement or debate may trigger repeated exchanges. Agents may inherit histories, share intermediate results, or build new contexts for later calls, changing opportunities for reuse~\cite{tokencake,tokendance}. Thus, even tasks with similar request counts and token lengths can differ in request readiness, batching opportunities, and memory demand (Fig.~\ref{fig:intro}). These differences arise from task structure, not merely scheduling decisions; request traces record the realized work, not the collaboration rules generating it.

Collaboration shapes serving demand in two distinct ways. \emph{Work generation} determines which model calls, inputs, and token lengths emerge as agents coordinate and exchange results. \emph{Work exposure} determines when these realized operations become eligible for execution under their dependencies. Legal changes to release constraints can alter batching opportunities and state lifetimes even when the recorded work remains fixed. These effects propagate through execution shapes, weight reuse, and memory demand to complete-task latency and deadline attainment. Understanding this chain requires both native workload realizations that capture collaboration-generated work and controlled comparisons that isolate the effects of its legal exposure to serving.

Existing work provides complementary perspectives. Multi-agent benchmarks evaluate collaboration effectiveness~\cite{masbench}, while workflow-aware serving optimizes particular mechanisms, such as round-based KV sharing~\cite{tokendance} and dependency-aware scheduling~\cite{maestro,tokencake,topas}. XPerf profiles agentic traces, reconstructs execution graphs, and synthesizes replay workloads by varying arrival patterns~\cite{xperf}; AgentSysBench characterizes complete applications~\cite{agentsysbench}, while AgentPerfBench profiles inference and hardware costs~\cite{agentperfbench}. MAServeBench instead makes documented collaboration patterns reusable and configurable through composable stages, allowing systems and architecture researchers to construct varied native serving workloads. Semantic Traces then enable legal release comparisons at fixed realized work.

We present \textbf{MAServeBench}, a benchmark that generates native workloads from configurable collaboration and context rules with identical task inputs and tools. Hierarchical Semantic Traces preserve each realization for legal release and resource comparisons. Task-linked measurements connect work generation and exposure to serving batches, hardware costs, and complete-task SLOs (Figure~\ref{fig:intro}(c)). \mbox{We make three contributions:}

\begin{itemize}

\item \textbf{Composable MAS workload construction.} We introduce reusable stages capturing agent coordination and context exchange while respecting state dependencies. They compose into MAS workflows whose collaboration rules vary independently of task bindings.

\item \textbf{Controlled work generation and exposure.} Native realizations generate collaboration-dependent work; hierarchical Semantic Traces preserve realized operations and typed dependencies, enabling controlled fixed-work replay on serving backends and providing workloads for architecture simulators.

\item \textbf{Task-linked hardware evaluation.} Functional and replay checks validate the benchmark. GPU/NPU studies connect MAS execution shapes to compute and memory demand, while controlled serving experiments reveal their implications for complete-task SLOs.

\end{itemize}

\section{Motivation and Benchmark Requirements}
\label{sec:requirements}

\subsection{Why Request-Level Workloads Are Insufficient}

\noindent\textbf{Work generation.} Arrival times, input/output lengths, and concurrency describe serving requests, but not the collaboration rules that produce them. Consider a research task in which independent agents collect evidence, a writer integrates their findings, and a reviewer may request revisions. Changing researcher count, revision rounds, or artifact delivery alters calls, contexts, and token work generated, even with shared task bindings and identical root arrivals.

\noindent\textbf{Work exposure.} At fixed realized work, a controlled comparison must identify legal release changes that preserve task semantics. The writer consumes the researchers' artifacts, while independent research calls can proceed concurrently. An added barrier may serialize these calls only if it preserves all artifact and state prerequisites and introduces no dependency cycle. Such ordering changes batching opportunities and state lifetimes without altering the recorded work. Selecting only part of a research artifact, by contrast, changes the writer's resolved input and requires a new native realization rather than a fixed-work replay.

\noindent\textbf{Hardware characterization.}
Attributing hardware costs to collaboration requires connecting task and stage identities to shared batches, operators, and resources, then locating their effects along task completion paths. Backend scheduling, caching, and placement determine how legally exposed work runs; these system choices must remain distinct from collaboration dependencies.

\subsection{Requirements for MAS Hardware Evaluation}

These gaps motivate four requirements for workload representation, controlled execution, and task-level measurement.

\noindent\textbf{R1: Representative Workloads.} A common workload interface must preserve essential participants, coordination, and context delivery across diverse documented MAS workflows and meaningful tasks.

\noindent\textbf{R2: Controllable Collaboration.} The benchmark must distinguish native changes to generated work from fixed-work comparisons that vary legal release constraints while preserving recorded requests and their required dependencies.

\noindent\textbf{R3: Reproducible Execution.}
Realized model, tool, transformation, and state operations must retain resolved inputs, artifacts, output contracts, and typed dependencies.
Replay must preserve this work while leaving target scheduling, batching, and placement decisions free.

\noindent\textbf{R4: Hardware Attribution.}
Measurements must connect task and stage progress to request readiness, batches, operators, and resources.
They must distinguish execution opportunities from hardware costs and relate both to complete-task latency and deadline-compliant service.

\section{Hierarchical Collaboration Workload Abstraction}
\label{sec:workload}

\begin{figure*}[t]
    \centering
    \includegraphics[width=\textwidth]{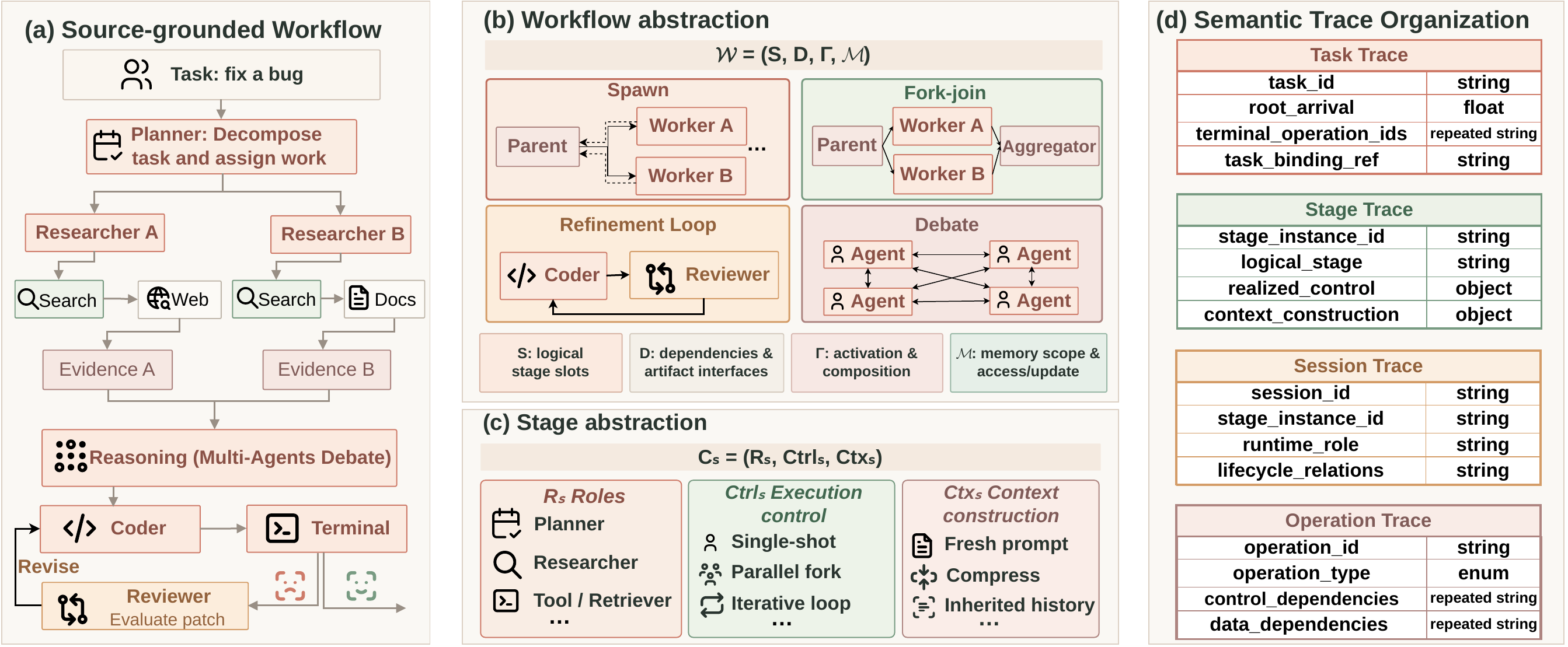}
    \vspace{-20pt}
    \caption{\textbf{MAServeBench hierarchical workload abstraction.}
    (a) Example multi-agent debugging workflow.
    (b) Four reusable collaboration templates within the workflow model.
    (c) Unified stage specification of participants, execution control, and context construction.
    (d) Hierarchical Semantic Trace records at task, stage, session, and operation levels.}
    \label{fig:masbench_method}
\end{figure*}

MAServeBench separates a \emph{logical workload} specifying collaboration rules from a \emph{realized execution} of concrete operations and dependencies.
Figure~\ref{fig:masbench_method} illustrates this abstraction; \S\,\ref{sec:benchmark} describes its realization and replay.

\subsection{Hierarchical Workload Model}
\label{sec:workload-hierarchy}

We represent an end-to-end task as
\begin{benchequation}
\mathcal{W}=(S,D,\Gamma,\mathcal{M}),
\qquad s\in S\mapsto\mathcal{C}_s,
\label{eq:workflow}
\end{benchequation}
where $S$ contains logical collaboration-stage slots and $D$ defines required inter-stage dependencies and artifact interfaces.
Workflow-level $\Gamma$ governs stage activation and composition through sequence, choice, parallel composition and join, bounded repetition, and handoff.
Logical memory $\mathcal{M}$ defines namespaces, sharing scopes, lifetimes, and access/update rules.

For an analysis--synthesis workflow, $S$ names stages, $D$ identifies required artifacts, $\Gamma$ composes analysis and synthesis, and $\mathcal{M}$ scopes their histories. Each stage has explicit input/output interfaces and a local specification $\mathcal{C}_s$. Repeated activations produce distinct stage instances and runtime sessions.

\subsection{Collaboration Stage Semantics}
\label{sec:stage-semantics}

Each stage specifies \emph{who participates}, \emph{how they coordinate}, and \emph{what information they receive}:
\begin{benchequation}
\mathcal{C}_s =
(R_s,\mathrm{Ctrl}_s,\mathrm{Ctx}_s).
\label{eq:stage}
\end{benchequation}

\noindent\textbf{Participants (\(R_s\)).}
\(R_s\) specifies participants as logical roles or agent/session slots with input/output interfaces. Roles may be instantiated zero, one, or multiple times according to the stage configuration and control decisions.

\noindent\textbf{Execution control (\(\mathrm{Ctrl}_s\)).}
\(\mathrm{Ctrl}_s\) governs local activation, concurrency, synchronization, iteration, handoff, and termination through spawn, fork, join, repeat, and finish. Unlike workflow-level composition $\Gamma$, these rules coordinate participants within a stage. They constrain operation readiness but leave batching, scheduling, and resource management to the serving backend.

\noindent\textbf{Context construction (\(\mathrm{Ctx}_s\)).}
\(\mathrm{Ctx}_s\) selects inherited or fresh history and delivered artifacts or peer outputs, then specifies their ordering and transformation into model inputs.
These choices alter generated token work and the opportunities for prefix/KV reuse.

\noindent\textbf{Logical memory and state.}
The workload-level $\mathcal{M}$ distinguishes private histories, task-shared versioned state, and externally persistent memory when supported.
It defines state scope and access rules, while $\mathrm{Ctx}_s$ selects information for each operation.
Versioned reads and writes induce state dependencies; retrieval and compression are explicit operations.
Logical state semantics remain distinct from physical KV allocation and eviction.

\subsection{Reusable Collaboration Templates}
\label{sec:templates}

The templates share the \(R_s/\mathrm{Ctrl}_s/\mathrm{Ctx}_s\) interface:

\noindent\textbf{Spawn.}
A parent activates children with inherited or fresh context and configured return and completion rules.

\noindent\textbf{Fork--Join.}
Independent branches join and aggregate outputs under configurable width and synchronization rules.

\noindent\textbf{Refinement Loop.}
An evaluator provides feedback on an intermediate output, triggering revision until acceptance or a configured round limit.

\noindent\textbf{Debate.}
Peers exchange outputs or critiques over bounded rounds, revising responses under configured topology, synchronization, and context rules.

Workflows compose these templates and atomic model, tool, or transformation stages through $\Gamma$.

\subsection{From Logical Workloads to Realized Execution}
\label{sec:realized-execution}

Each run \(r\) realizes a concrete operation graph:
\begin{benchequation}
G^{(r)} =
\left(
O^{(r)},
E_{\mathrm{ctrl}}^{(r)}
\cup E_{\mathrm{data}}^{(r)}
\cup E_{\mathrm{state}}^{(r)}
\right),
\label{eq:realized}
\end{benchequation}
where \(O^{(r)}\) contains realized model, tool, transformation, and state operations.
Activation, branching, and iteration determine which operations occur in each run.
The three edge sets encode control prerequisites, consumed artifacts, and versioned state dependencies, respectively.

Each operation retains its hierarchical identity:
\begin{benchequation}
h(o) =
(\text{task},\text{stage instance},\text{session}).
\label{eq:hierarchy-map}
\end{benchequation}
The session may be absent for transformation or state operations; stage-instance identities and iteration metadata distinguish repeated activations.
This mapping links serving and operator observations to task and stage progress.

\S\,\ref{sec:benchmark} records this graph with resolved payloads, output contracts, and external delays in a Semantic Trace.

\section{Benchmark Design}
\label{sec:benchmark}

MAServeBench combines native generation, hierarchical Semantic Traces, and target-side measurement to turn the workload abstraction into controlled serving experiments.

\subsection{Benchmark Suite and Workload Construction}
\label{sec:Construction}

A \emph{task binding} supplies inputs, role instructions, tools, and evaluation criteria independently of collaboration rules. The templates in \S\,\ref{sec:workload} form standalone and composed workflows with configurable control and artifact delivery under shared bindings.
\emph{Native realization} executes collaboration rules and tools to generate concrete operations, inputs, and dependencies. \emph{Fixed-work replay} preserves those operations, resolved inputs, output contracts, and required dependencies while varying legal release constraints or serving resources.

In a four-call workflow, two analysts feed synthesis before verification. An optional analyst-to-analyst barrier changes release order while preserving required dependencies. Selecting each analyst's answer field instead of its full JSON artifact changes synthesis input through a recorded CPU transformation, producing a new native realization (\S\,\ref{sec:eval-control}).

\subsection{Semantic Trace}
\label{sec:trace-execution}

\edef\masRestoreSchemaFloatSpacing{%
  \floatsep=\the\floatsep\relax
  \textfloatsep=\the\textfloatsep\relax
}
\setlength{\floatsep}{6pt}
\setlength{\textfloatsep}{8pt}

Tables~\ref{tab:trace_hierarchy} and~\ref{tab:trace_operation} define the manifest and operation schemas. A versioned \emph{Semantic Trace} records realized operations with typed dependencies, resolved inputs and context references, generation settings, optional native output token IDs and output-length targets, tool results and measured elapsed times, and controlled external delays. During replay, tool operations return recorded results after their recorded durations without invoking external services; downstream operations remain subject to the same dependencies. Payloads and artifacts reside in a content-addressed store, preserving inputs consumed by downstream operations. Versioned state records retain read/write provenance, visibility, and resolved memory-derived inputs. Replay starts from reproducible initial state. Persistent-store replay depends on reproducible snapshots and restoration hooks.

\noindent\textbf{Trace and comparison contracts.}
A \emph{trace} fixes realized work, resolved inputs, output contracts, and required dependencies. A \emph{scenario} binds a trace set to root arrivals, replay/release rules, model-input contracts, initial cache/state, measurement settings, and task SLOs. A \emph{system configuration} specifies the backend, hardware, and scheduling, batching, caching, placement, or resource policies.

A legal, stage-annotated barrier may serialize independent operations, preserving required dependencies and rejecting cycles. Resource comparisons retain release rules and root arrivals while varying a backend setting such as the KV~budget.
Cross-system comparisons fix the trace and scenario while changing the system configuration. Backend routing and placement remain separate from logical task and session identities and collaboration dependencies.

\subsection{Execution and Hardware Measurement}
\label{sec:benchmark-metrics}

Root tasks enter at scenario arrivals. Operations become ready after their control, data, and state predecessors complete and external delays elapse; the backend handles admission, scheduling, and execution.
\emph{Length-locked replay} fixes recorded inputs and output lengths while the model generates its own tokens. \emph{Token-locked replay} replaces each sampled token with its native recorded ID~\cite{xperf,agentreplay}. Both perform prefill and autoregressive decoding, retaining downstream artifacts and required dependencies. Token control is verified on CUDA vLLM Model Runner V2 with dense Qwen3 (Appendix~\ref{app:token-replay}).

\noindent\textbf{Target-side measurement interface.}
Execution keys or aligned timestamps trace task, stage, and operation IDs through serving requests to scheduler steps, shared batches, and native capture windows. Records retain readiness, submission, first scheduling, completion, batch composition, operator shapes, and resources. Shared kernels retain participating-request sets, linking operator observations back to the tasks and stages that exposed their work.

\begin{table}[t] \centering \scriptsize \setlength{\tabcolsep}{4pt} \renewcommand{\arraystretch}{1.05} \caption{MAServeBench Semantic Trace manifest schema. Brackets [] denote repeated records.} \label{tab:trace_hierarchy} \begin{tabular}{|>{\raggedright\arraybackslash}p{0.27\linewidth}|>{\raggedright\arraybackslash}p{0.29\linewidth}|>{\raggedright\arraybackslash}p{0.33\linewidth}|} \hline \textbf{Field Name} & \textbf{Data Type} & \textbf{Description} \\ \hline metadata & TraceMetadata & Version and identity \\ \hline tasks & TaskRecord[] & End-to-end tasks \\ \hline stages & StageInstance[] & Realized stages \\ \hline sessions & AgentSession[] & Runtime participants \\ \hline artifacts & ArtifactRecord[] & Delivered artifacts \\ \hline shared\_state & StateRecord[] & Versioned state \\ \hline operations\_file & string & Operation stream \\ \hline \end{tabular} 
\par\vspace{-10pt}
\end{table} 
\begin{table}[t] \centering \scriptsize \setlength{\tabcolsep}{4pt} \renewcommand{\arraystretch}{1.00} \caption{MAServeBench semantic operation schema.} \label{tab:trace_operation} \begin{tabular}{|>{\raggedright\arraybackslash}p{0.34\linewidth}|>{\raggedright\arraybackslash}p{0.23\linewidth}|>{\raggedright\arraybackslash}p{0.32\linewidth}|} \hline \textbf{Field Name} & \textbf{Data Type} & \textbf{Description} \\ \hline operation\_id & string & Unique operation ID \\ \hline operation\_type & enum(OpType) & LLM, tool, transform, state \\ \hline task\_id & string & Parent task \\ \hline stage\_instance\_id & string & Parent stage instance \\ \hline session\_id & optional string & Parent agent/session \\ \hline iteration & uint32 & Iteration index \\ \hline control\_\allowbreak dependencies & repeated string & Control prerequisites \\ \hline data\_dependencies & repeated string & Artifact prerequisites \\ \hline state\_dependencies & repeated string & State prerequisites \\ \hline controlled\_\allowbreak external\_\allowbreak delay & float & Controlled external delay \\ \hline llm/\allowbreak tool/\allowbreak transform/\allowbreak state & optional object & Typed payload \\ \hline source\_observations & map & Source diagnostics \\ \hline \end{tabular} \end{table}

\FloatBarrier

Unprofiled runs measure application latency; separate native captures measure kernel durations and physical accesses within sampled windows. Allocated KV occupancy measures reserved serving state. Matched shape and working-set controls characterize operator costs on each hardware/implementation stack (Appendices~\ref{app:measurement},~\ref{app:compute}--\ref{app:memory}).

\noindent\textbf{Task-level metrics.} Root latency spans arrival to terminal completion; stage time spans first-operation readiness to stage completion. Attempt records cover completed, failed, and incomplete roots plus consumed work. Completion-path accounting separates pre-submission (ready to submitted), backend waiting (submitted to first scheduling), and scheduled residence (first scheduling to completion).

For external arrivals, per-class deadlines are fixed before configuration comparisons. SLO attainment counts deadline-compliant completions among all attempted roots, and task goodput divides these completions by the declared observation period. Appendix~\ref{app:measurement} specifies observation windows and metric contracts.
\par
\masRestoreSchemaFloatSpacing

\section{Validating MAServeBench}
\label{sec:validity}

We validate MAServeBench at three levels: correspondence to documented collaboration patterns, functional comparability on real tasks, and fidelity of recorded-work replay.

\subsection{Source-Grounded Abstraction Validation}
\label{sec:validity-workflows}

\noindent\textbf{Structural coverage.} Twelve documented workflows cover all four local templates (Table~\ref{tab:workflow-grounding}); five compose multiple templates, while MetaGPT uses sequential handoff through $\Gamma$. The executable suite exercises seven structures across 30 task bindings, yielding 210 native realizations. A separate 21-instance fixed bank, with three bindings per structure, characterizes operation-graph depth, parallel width, and typed dependencies (Appendix~\ref{app:source-mappings}). An eight-binding GPT Researcher comparison further examines source-to-realization dependency geometry (Appendix~\ref{app:source-serving-correspondence}).

\begin{table*}[t]
\caption{\textbf{Source-grounded collaboration mappings.} Twelve workflows: templates $\mathcal{T}$ (first line) and participants $\mathcal{P}$ (second line). MetaGPT uses $\Gamma$ for sequential handoff ($\mathcal{T}=\varnothing$).}
\label{tab:workflow-grounding}
\centering
\scriptsize
\setlength{\tabcolsep}{4pt}
\renewcommand{\arraystretch}{1.05}
\begin{tabularx}{\textwidth}{
    @{}
    >{\raggedright\arraybackslash}p{.145\textwidth}
    >{\raggedright\arraybackslash}X
    @{\hspace{.018\textwidth}}
    >{\raggedright\arraybackslash}p{.175\textwidth}
    >{\raggedright\arraybackslash}X
    @{}
}
\toprule
\textbf{Workflow} & \textbf{Templates / participants} &
\textbf{Workflow} & \textbf{Templates / participants} \\
\midrule
\textbf{MetaGPT startup}\par\cite{metagpt} &
$\Gamma$: Sequential Handoff\par
Product Manager, Architect, Project Manager, Engineer &
\textbf{OWL Workforce}\par\cite{owl_workforce} &
\emph{Spawn}\par
Planner, Coordinator, Workers \\
\addlinespace[1pt]
\textbf{ChatDev Default}\par\cite{chatdev} &
\emph{Refinement Loop}\par
Programmer, Code Reviewer, Software Test Engineer &
\textbf{TradingAgents}\par\cite{tradingagents} &
\emph{Debate}\par
Bull and Bear Researchers, Risk Analysts \\
\addlinespace[1pt]
\textbf{Magentic-One}\par\cite{magenticone} &
\emph{Spawn}\par
Orchestrator, Selected Workers &
\textbf{Virtual Lab}\par\cite{virtual_lab} &
\emph{Fork--Join, Debate}\par
Meeting Leads and Members, Merge Participants \\
\addlinespace[1pt]
\textbf{MindSearch}\par\cite{mindsearch} &
\emph{Spawn}\par
Planner, Searcher &
\textbf{Claude Code PR review}\par\cite{claude_code_agent_teams} &
\emph{Fork--Join}\par
Lead, Security, Performance, Test Reviewers \\
\addlinespace[1pt]
\textbf{GPT Researcher}\par\cite{gpt_researcher_multiagent} &
\emph{Fork--Join, Refinement Loop}\par
Editor, Researcher, Reviewer, Reviser, Writer &
\textbf{DeerFlow literature review}\par\cite{deerflow_slr} &
\emph{Spawn, Fork--Join}\par
Lead, Extraction Subagents \\
\addlinespace[1pt]
\textbf{Open Deep Research}\par\cite{open_deep_research} &
\emph{Spawn, Fork--Join}\par
Supervisor, Researchers &
\textbf{Agyn issue resolution}\par\cite{agyn} &
\emph{Spawn, Refinement Loop}\par
Manager, Researcher, Engineer, Reviewer \\
\bottomrule
\end{tabularx}
\end{table*}

\noindent\textbf{Functional validity.} Seven paired native comparisons cover six frameworks and four task sets: SWE-bench Verified, GAIA, WebArena, and DeepResearch Bench. Each pair uses matched task instances, Qwen3.5-9B settings, and evaluation procedures; both the reference workflow and its MAServeBench realization execute natively, allowing generated outputs to determine subsequent work. Quality differences of 1.2--3.8 percentage points on success/accuracy and at most 0.019 on RACE indicate broadly comparable task outcomes. Comparable task outcomes can still require different model/tool calls and token work, making serving-demand measurement necessary (Appendix~\ref{app:functional-validation}).

\subsection{Preserving Serving-Relevant Semantics}
\label{sec:validity-semantics}

\noindent\textbf{Semantic preservation.} The mappings retain source dependency order, artifact delivery, and context construction~\cite{tradingagents,virtual_lab}. Research workflows retain review ordering, isolated contexts, and explicit compression~\cite{gpt_researcher_multiagent,open_deep_research}, while Claude Code task/message prerequisites become versioned state dependencies~\cite{claude_code_agent_teams}.

Across three paired native evaluations, request-input medians differ by at most 2.3\%, with the largest input P90 shift a 7.1\% reduction for Magentic-One/WebArena. Across eight GPT Researcher bindings, both arms share medians in model calls (30), graph depth (20), antichain width (3), and peak client in-flight (3); three client requests overlap for 23\% of root lifetime on average in both arms, with larger prompts later in execution. Appendices~\ref{app:functional-validation} and~\ref{app:source-serving-correspondence} detail these paired native experiments.

\noindent\textbf{Execution conformance.} Recorded-work checks verify input/output counts for 1,032 completed requests and 186 required predecessor-release constraints within a 1\,ms tolerance. Native-to-replay comparisons on Qwen3-8B/910C cover Fork--Join and Fork--Join with Refinement at one or 16 concurrent roots, with prefix caching enabled or disabled and cold-cache starts (Appendix~\ref{app:execution-validation}). All 204 paired roots preserve resolved payloads, token counts, operations, and dependencies exactly, with no early predecessor release; replay-to-native root-latency differences range from $-1.85\%$ to $+2.88\%$. Work preservation, correct release, and timing fidelity support the controlled comparisons.
\section{Evaluation}
\label{sec:eval}
\raggedbottom
\clubpenalty=10000
\widowpenalty=10000
\makeatletter
\edef\masRestoreParagraphBreaks{%
  \brokenpenalty=\the\brokenpenalty\relax
  \@clubpenalty=\the\@clubpenalty\relax
  \textfloatsep=\the\textfloatsep\relax
  \floatsep=\the\floatsep\relax
}
\brokenpenalty=1000
\@clubpenalty=10000
\makeatother
\setlength{\textfloatsep}{10pt plus 2pt minus 2pt}
\setlength{\floatsep}{6pt plus 1pt minus 1pt}
\setcounter{topnumber}{4}
\setcounter{dbltopnumber}{4}
\renewcommand{\topfraction}{.92}
\renewcommand{\dbltopfraction}{.92}
\renewcommand{\textfraction}{.05}

Source-grounded native comparisons assess workflow correspondence; controlled MAS cohorts expose dependency, decoding, and KV regimes, while matched GPU/NPU probes measure hardware sensitivity. Fixed-work interventions isolate release, capacity, and co-execution effects on task latency and goodput. Additional protocols and results appear in the appendix. 

\subsection{Experimental Setup}
\label{sec:eval-setup}

We use vLLM and vLLM-Ascend on the GPU/NPU platforms in Table~\ref{tab:eval-platforms}, with BF16, tensor parallelism one, and one exposed accelerator per serving instance. Qwen3.5 combines recurrent linear and full attention; Qwen3-8B~\cite{yang2025qwen3technicalreport} provides full-attention workloads.

Seven structures cover the four templates, two compositions (Spawn--Fork--Join--Refinement and Fork--Join--Debate--Refinement), and a research--review--writing Sequence. Thirty shared task bindings retain generated artifacts and natural output lengths across code, reasoning, finance, evidence processing, and science.

Three-block native runs use 32/64/128 users, each issuing one ordinary call or executing a MAS workflow. Three paired Qwen3-8B/910C blocks fix work and arrivals for release/resource interventions. The GPU composition study limits requests to 64. Fixed-work serving evaluations use length-locked replay unless stated otherwise.

\begin{table}[t]
\centering\small

\caption{\textbf{Application and operator coverage.} 910C applications use one logical chip per dual-chip card; operator probes are separate from application runs.}
\label{tab:eval-platforms}
\fontsize{8}{9.5}\selectfont
\setlength{\tabcolsep}{3pt}
\begin{tabularx}{\columnwidth}{@{}p{.17\columnwidth}>{\raggedright\arraybackslash}X>{\raggedright\arraybackslash}X@{}}
\toprule
Platform & Application workload & Matched operator \\
\midrule
A800 & Qwen3.5-4B/9B/27B & Qwen3.5-9B projections \\
PRO 5000 & Qwen3.5-9B & Qwen3.5-9B projections \\
910C & Qwen3.5-9B; Qwen3-8B & Matrix, normalization, gather; full attention \\
950DT & Qwen3.5-9B captures & Matrix, normalization, gather; Qwen3-8B attention \\
\bottomrule
\end{tabularx}
\end{table}

\subsection{Generated Demand and Controlled Work~Exposure}
\label{sec:eval-control}

\begin{figure}[t]
\centering
\includegraphics[width=\columnwidth]{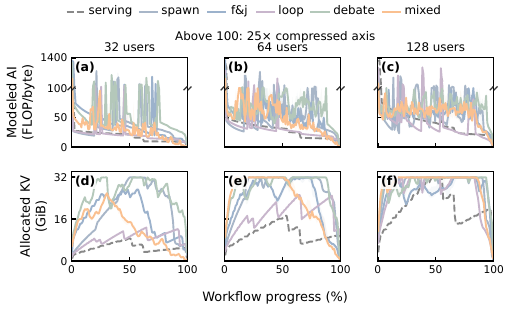}
\vspace{-20pt}
\caption{\textbf{Demand over workflow progress.} Native Qwen3-8B/910C, 32/64/128 users; three-block medians/ranges. (a)--(c) Modeled AI (FLOP/byte), with marked $25\times$ axis compression above 100. (d)--(f) Allocated KV (GiB).}
\label{fig:eval-demand}
\end{figure}

\noindent\textbf{Natural workload demand.} Spawn and Fork--Join expose parallel calls, whereas Loop advances along dependencies. Their native demand unfolds differently (Figure~\ref{fig:eval-demand}): at 64 users, Spawn has $1.70\times$ ordinary serving's aggregate modeled arithmetic intensity, while its longer execution (580 versus 109 seconds) contributes to a $13.85\times$ KV memory--time integral. Loop has comparable intensity but longer state residency.

\begin{figure}[t]
\centering
\includegraphics[width=\columnwidth]{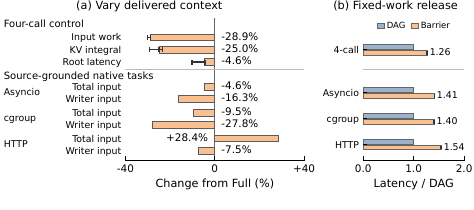}
\vspace{-20pt}
\caption{\textbf{Native generation versus fixed-work release.} (a) Selected changes from Full: four-call controls (three-block medians/ranges) and GPT Researcher (one native pair per binding; writer input denotes the first call). (b) Barrier/DAG root-latency ratios over three paired replay blocks.}
\label{fig:eval-release}
\end{figure}

\noindent\textbf{Native context delivery.} In the four-call control with matched analyst inputs, Selected reduces input work by 29\% and KV memory--time by 25\%, but root latency by only 5\% (paired-block medians; Figure~\ref{fig:eval-release}(a)).

\noindent\textbf{Source-workflow generation.} Three GPT Researcher bindings realize planning, parallel research/refinement, and writing/feedback. Full delivers reviewed stage states to the writer; Selected delivers drafts. For HTTP, native Selected reduces first-writer input by 7.5\% yet increases total input by 28.4\%; an additional section revision and writing/checking pass expand LLM calls from 25 to 30 and graph depth from 15 to 20. Editing the frozen Full graph cannot generate these additional revision/writing calls; a new native realization is required (Figure~\ref{fig:eval-release}(a); Appendix~\ref{app:legal-control}).

\noindent\textbf{Legal control at fixed work.} We freeze the four-call Full traces and vary only legal $\mathrm{Ctrl}_s$ constraints. Barrier adds analyst ordering, reducing scheduled width from 24 to 12 and increasing mean root latency by 26\% (Figure~\ref{fig:eval-release}(b)); Selected replays show 28\% (Appendix~\ref{app:legal-control}). For three frozen GPT Researcher Full traces (83 calls), ordering research branches preserves paired arrivals, reduces scheduled width from 9 to 3, and increases root latency by 40--54\% (Figure~\ref{fig:eval-release}(b)). \textbf{These fixed-work comparisons isolate release effects from generated demand.}

\benchfinding{1: Collaboration controls generated work and its exposure}{\textbf{Native generation:} Changing context delivery can reduce local input while increasing total realized work. \textbf{Fixed-work exposure:} Legal ordering reduces scheduled width and raises root latency by 26--54\%.}

\subsection{Compute Demand}
\label{sec:eval-hardware}

\begin{figure}[t]
\centering
\includegraphics[width=\columnwidth]{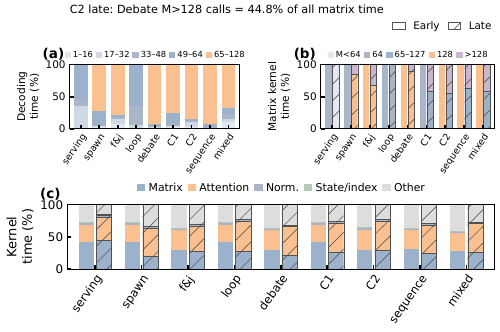}
\vspace{-20pt}
\caption{\textbf{Native execution shapes and kernel mix.} Qwen3-8B/910C, 64 users, 128-request limit; Spawn has four children. (a) Time-weighted decoding batches, three-block average. (b) Matrix time by tensor-row $M$. (c) Kernel-time mix. (b,c) One capture per workflow/window; Early/Late: plain/hatched. C1/C2 follow the composition order in \S\,\ref{sec:eval-setup}.}
\label{fig:eval-execution}
\end{figure}

\noindent\textbf{Native execution shapes.} At 64 root users, parallel workflows decode mostly above 64 requests, while ordinary serving and Loop remain within 64 (Figure~\ref{fig:eval-execution}(a)). Spawn's modal matrix rows $M$ shift from 64 early to 128 late, versus 64 for Loop (b). Growing contexts raise mixed-cohort attention's kernel-time share from 28.3\% to 44.6\% (c). Semantic Trace attribution links C2's late $M>128$ calls to Debate, contributing 44.8\% of that capture's matrix kernel time. \textbf{Stage-linked traces identify the operator shapes exposed by collaboration stages. We use these workload-derived shapes to construct matched probes that vary batch geometry, weight locality, and execution paths.}

\begin{table}[!b]
\centering
\small
\setlength{\tabcolsep}{3pt}
\renewcommand{\arraystretch}{0.95}
\caption{\textbf{GPU/NPU architectural context.} Specifications follow NVIDIA~\cite{nvidia_ampere_arch,nvidia_rtx_blackwell_arch,nvidia_pro5000_specs} and Huawei~\cite{huawei_950_arch}. A800 is the 80\,GB SXM4 variant; NPU vector-access granularity is vendor-reported.}
\label{tab:accelerator-architecture}
\fontsize{8}{9.5}\selectfont
\begin{tabularx}{\columnwidth}{@{}p{.21\columnwidth}>{\raggedright\arraybackslash}X>{\raggedright\arraybackslash}X@{}}
\toprule
\multicolumn{3}{@{}l}{\textbf{GPU}} \\
\textbf{Feature} & \textbf{A800 (Ampere)} & \textbf{PRO 5000 (Blackwell)} \\
\midrule
Architecture & GA100, \texttt{sm\_80} & GB202, \texttt{sm\_120} \\
BF16 matrix & 3rd-gen Tensor Cores & 5th-gen Tensor Cores \\
L2 cache & 40\,MiB & 96\,MiB \\
Memory & 80\,GB HBM2e & 48\,GB GDDR7 \\
Bandwidth & 2.04\,TB/s & 1.34\,TB/s \\
\midrule
\multicolumn{3}{@{}l}{\textbf{NPU}} \\
\textbf{Feature} & \textbf{Ascend 910C} & \textbf{Ascend 950DT} \\
\midrule
Execution & Cube matrix; SIMD vector & Cube; expanded vector, SIMD/SIMT \\
Vector access & Reported 512-byte granularity & Reported 128-byte granularity \\
\bottomrule
\end{tabularx}
\end{table}

\begin{figure}[t]
\centering
\includegraphics[width=\columnwidth]{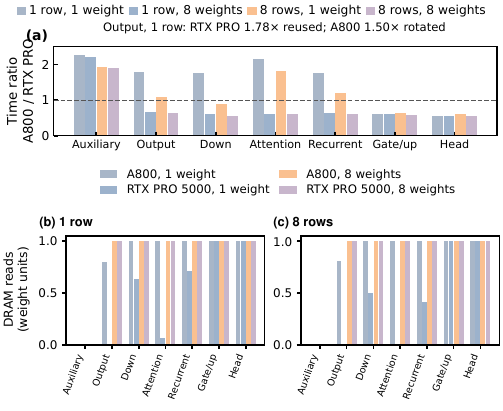}
\vspace{-20pt}
\caption{\textbf{GPU projection response to geometry and weight locality.} (a) All 28 timing ratios, A800/RTX PRO 5000; higher favors RTX PRO 5000. (b,c) One/eight-row physical DRAM reads normalized by one BF16 weight, from independent counter passes.}
\label{fig:eval-gpu-compute}
\label{fig:eval-gpu-memory}
\end{figure}

\noindent\textbf{GPU: Geometry and locality.} Native Qwen3.5-9B serving exposes low-batch final-drain phases (Appendix~E), while repeated projection shapes across layers use distinct weight addresses. We probe geometry and locality in 28 cases spanning seven model-derived $N,K$ families, one/eight rows, and one/eight weight addresses (Figure~\ref{fig:eval-gpu-compute}(a)). For the one-row $4096\times4096$ output projection, PRO 5000 is $1.78\times$ faster with weight reuse, whereas A800 is $1.50\times$ faster under eight-weight rotation. L2 capacity and bandwidth differences (Table~\ref{tab:accelerator-architecture}) help explain this reversal: independent counters record no external reads on PRO 5000 versus 25.6\,MiB on A800 for one reused 32\,MiB weight. Rotation expands the working set to 256\,MiB, exceeding both caches, with roughly 32\,MiB read per call (b,c). This sensitivity also informs multi-model MAS serving on shared accelerators, where interleaved weights enlarge active working sets.

\begin{figure}[t]
\centering
\includegraphics[width=\columnwidth]{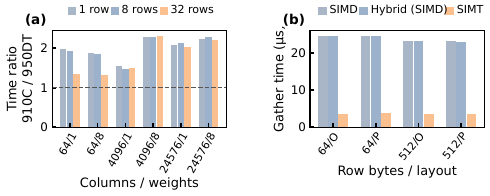}
\vspace{-20pt}
\caption{\textbf{NPU compute response.} (a) 910C/950DT Memory-pass ratios for 18 BF16 matrix cases, $K=4096$. (b) Twelve 950DT Gather implementations; binary inspection verifies SIMT and identifies Hybrid as SIMD. O/P: ordered/permuted layouts.}
\label{fig:eval-npu-compute}
\end{figure}

\noindent\textbf{NPU: Shape and execution paths.} Three matrix reference families cover 54.9\% of sampled 910C hybrid-model matrix time by $N,K$. The 18 matched cases vary rows and weight reuse (Figure~\ref{fig:eval-npu-compute}(a)); at 32 rows and $4096\times4096$, 950DT's native Memory pass is $1.49\times$ faster with one reused weight and $2.32\times$ with eight rotating weights. Native state/index activity motivates Gather probes: across 12 implementations on 950DT, binary-verified SIMT outperforms SIMD by $6.47$--$6.90\times$, while Hybrid emits the same SIMD binary (b).

\benchfinding{2: Compute needs locality and flexible execution}{Native MAS stages expose changing batch and matrix shapes. \textbf{Weight locality reverses the GPU projection ranking} between reuse and rotation, while binary-verified SIMT Gather outperforms SIMD by $6.47$--$6.90\times$ on 950DT.}

\subsection{Memory Demand}
\label{sec:eval-capacity}

\begin{figure}[t]
\centering
\includegraphics[width=\columnwidth]{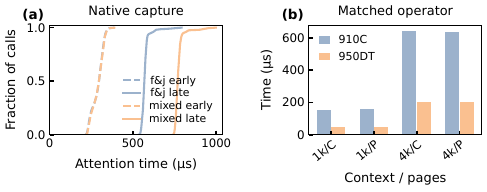}
\vspace{-20pt}
\caption{\textbf{NPU attention response to context geometry.} (a) Native Qwen3-8B/910C attention durations. (b) Matched 910C/950DT probes at batch 32 with 1k/4k contexts and contiguous/permuted page layouts (C/P).}
\label{fig:eval-npu-memory}
\end{figure}

\noindent\textbf{Context-dependent attention.} Collaboration-defined artifact delivery shapes downstream attention contexts. In native Qwen3-8B/910C mixed-cohort captures, median attention duration rises from 296 to 769\,$\mu$s between early and late execution (Figure~\ref{fig:eval-npu-memory}(a)). Eight application-derived batch-64 controls on 910C halve the exact context vectors at fixed memory backing, yielding full/half kernel-time ratios of $1.91$--$2.01\times$ (Table~F.1). Matched batch-32 probes compare 910C and 950DT across 1k/4k contexts and contiguous/permuted page layouts (b). Quadrupling context raises attention time roughly fourfold on both NPUs, while 950DT is $3.08$--$3.17\times$ faster at matched geometry. The 1k--4k range covers 95.8--98.7\% of time-weighted pure-decode context observations in the application controls (Appendix~F).

\begin{figure}[t]
\centering
\includegraphics[width=\columnwidth]{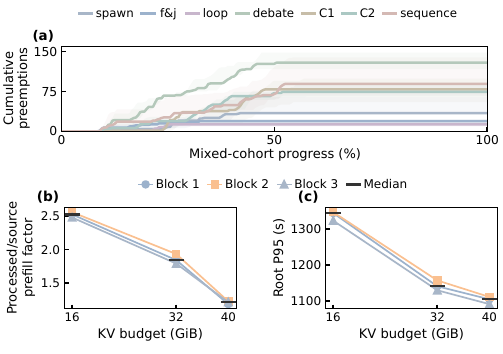}
\vspace{-20pt}
\caption{\textbf{KV pressure and task tails.} Qwen3-8B/910C, 64 users. (a) Native preemptions by victim workflow at 31.99\,GiB; three-block medians/ranges. (b,c) Fixed-work prefill factor and root P95. Connected points: paired blocks; black: medians.}
\label{fig:eval-state}
\label{fig:eval-budget}
\end{figure}

\noindent\textbf{KV pressure and task tails.} Parallel branches keep agent contexts live concurrently, increasing contention for KV residency. Across three native Qwen3-8B/910C mixed-cohort blocks, 1,372 preemptions amplify processed prefill to $1.76$--$1.84\times$ source input; Loop-only blocks show neither effect (Figure~\ref{fig:eval-state}(a)). All preemptions fall in time bins with at least 95\% mean KV allocation, linking eviction to costly recomputation of existing contexts. To isolate capacity, a 64-root fixed-work replay preserves arrivals, token work, and dependencies across 16/32/40\,GiB KV budgets. Greater capacity retains overlapping states, reducing preemptions, redundant prefill, and root P95 in every paired block (b,c). Native traces thus identify collaboration-shaped state pressure; fixed-work controls measure its latency cost, while \S\,\ref{sec:eval-slo} tests release and capacity against task deadlines.

\benchfinding{3: Retain overlapping KV states}{Parallel agents increase concurrent KV residency, turning preemption into costly repeated prefill under memory pressure. At fixed work, expanding KV from 16 to 40\,GiB reduces the \textbf{processed/source prefill factor by 52\%} and root P95 by 17.5\%.}

\subsection{From Stage Execution to Task Latency}
\label{sec:eval-interference}

Fixed-work Semantic Trace replay isolates how background timing changes co-batching, dependent-call residence, and root latency. We test this with three paired Qwen3-8B/910C blocks replaying nine Loop targets (three per binding) and 55 background roots (19 Loop, 18 Spawn, 18 Fork--Join). Interleaved A and clustered B reassign the same background traces to fixed global arrival slots; \textbf{recorded work, dependencies, and target arrivals remain identical}.

\begin{figure}[t]
\centering
\includegraphics[width=\columnwidth]{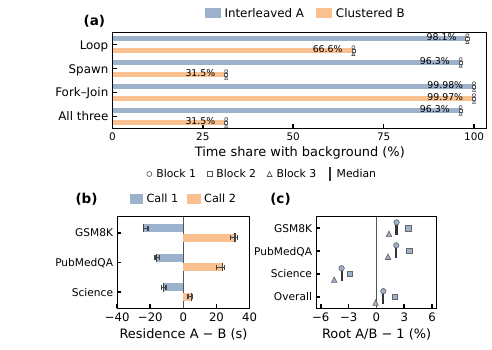}
\vspace{-20pt}
\caption{\textbf{Temporal mixing at fixed work.} (a) Time-weighted share of target-containing scheduled intervals co-batched with each background class or all three, 2--110\,s (bars: medians; markers: blocks). (b) First/second-call scheduled-residence differences ($A-B$; paired block means, median/range). (c) Root-latency changes.}
\label{fig:eval-temporal-mixing}
\end{figure}

\textbf{Actual co-batching changes substantially}: target-containing scheduled time overlapping all three background classes rises from 31.5\% (B) to 96.3\% (A) over 2--110\,s (Figure~\ref{fig:eval-temporal-mixing}(a)). \textbf{All 27 targets have shorter first-call but longer second-call scheduled residence under A} (b), as earlier first-call completion releases the dependent call into a different period of shared execution. The second-call penalty dominates for GSM8K (+2.17\%) and PubMedQA (+2.14\%), but not Science ($-3.75\%$), yielding a small overall change (+0.743\%, median of three within-block ratios; c). With zero KV preemptions and empty backend waiting queues, these differences arise mainly in scheduled residence (first scheduling to completion, including pauses).

\begin{figure}[t]
\centering
\includegraphics[width=\columnwidth]{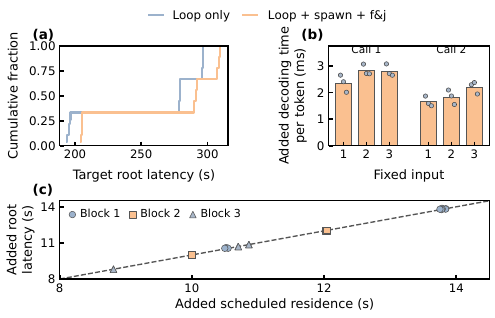}
\vspace{-20pt}
\caption{\textbf{Fixed Loop targets under near-matched background.} (a) Root latency. (b) Added client-observed decoding time per output token; bars/dots: mean/block means, two calls and three frozen inputs. (c) All 27 paired targets; the line denotes equal root-latency and scheduled-residence changes.}
\label{fig:eval-cotenancy}
\end{figure}

\noindent\textbf{Background composition and demand.} A separate near-matched comparison fixes the 9 Loop targets but replaces 2 of 55 Loop background roots with one Spawn and one Fork--Join (Figure~\ref{fig:eval-cotenancy}), adding six model calls and 4.5\%/4.8\% background input/output tokens. Target mean latency rises 3.9--5.0\% across blocks, mainly in scheduled residence (a,c); both calls show longer client-observed effective decoding time per token, including scheduling effects (b).

\subsection{From Hardware Efficiency to Task SLOs}
\label{sec:eval-slo}

We test how legal release and KV capacity affect task deadlines using a $2\times2$ DAG/Barrier $\times$ 16/40\,GiB comparison for Spawn--Fork--Join--Refinement on Qwen3-8B/910C. Barrier adds three ordering edges among four independent analyses. Across four arrival rates and three paired blocks, recorded work, required dependencies, and Poisson arrivals remain fixed. The shared 361.7-second deadline is $1.5\times$ an independent low-load DAG/40\,GiB reference P95. Task goodput counts timely roots, including drain completions, per arrival-window duration; output TPS uses drain-inclusive duration.

\begin{figure}[t]
\centering
\includegraphics[width=\columnwidth]{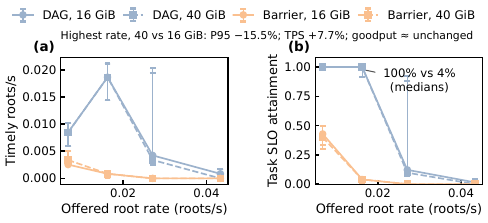}
\vspace{-20pt}
\caption{\textbf{Legal release under deadlines.} Fixed-work (a) arrival-window deadline goodput and (b) attainment. Colors: release rules; line styles: KV budgets. Points: three-block medians/ranges.}
\label{fig:eval-slo}
\end{figure}

\textbf{Release constraints affect task completion even without KV recomputation.} The first three rates incur no KV preemption, yet at the second rate DAG achieves 100\% median deadline attainment versus Barrier's 4\% under either KV budget (Figure~\ref{fig:eval-slo}).
\textbf{Improved hardware efficiency does not guarantee better task SLOs.} At the highest rate, 40\,GiB eliminates preemption observed at 16\,GiB in two blocks per release rule, reducing DAG root P95 by 15.5\% and increasing output TPS by 7.7\%. Nevertheless, deadline goodput remains near zero: these execution savings do not move enough complete-task latencies below the shared deadline. The DAG--Barrier advantage persists across tested deadlines (Figure~\ref{fig:app-deadline-sensitivity}). Separately, a higher ordinary-serving request limit (64 to 128) raises output TPS by 14.6\% but lowers timely-task rate by 22.5\%, reversing the preferred configuration (Appendix~\ref{app:request-limit}).

\benchfinding{4: Hardware gains must meet task deadlines}{Release ordering and KV capacity must be judged by timely task completion. At moderate load, DAG achieves 100\% versus Barrier's 4\% deadline attainment; at high load, extra KV raises TPS by 7.7\% without improving deadline goodput.}

\subsection{Discussion}
\label{sec:eval-platform}

\noindent\textbf{Reuse-aware compute and tiered KV memory.} The GPU reversal motivates SRAM/L2 retention for reusable weights and cache bypass with HBM streaming for rotating weights; Gather motivates programmable SIMT with access-aware path selection. A page-managed HBM pool could asynchronously extend KV into DRAM when migration and remote access cost less than saved prefill. High-Bandwidth Flash (HBF) could hold cold, read-mostly weights with low update traffic; 3D near-memory attention could benefit long contexts when saved data movement exceeds transfer overhead. These hypotheses must advance timely task completion; dependency hints and input-prefix reuse could guide retention and prefetch.

\noindent\textbf{Multi-replica and disaggregated serving.} Consider a production MAS whose agents invoke model APIs backed by replicas across machines or clusters. Prefix affinity can save prefill but concentrate demand on a contended replica, making routing and load balancing affect complete-task latency and SLOs even with fixed dependencies. MAServeBench separates logical collaboration from physical placement; an API-facing replay adapter could preserve recorded work and arrivals while exposing per-replica routing and queueing, including delays before submission.
At the phase level, systems such as Mooncake~\cite{mooncake} separate prefill and decode~\cite{distserve,splitwise}. Parallel MAS stages create prefill bursts, while refinement and debate repeatedly release dependent calls. Phase specialization must offset KV/state-transfer costs and cross-pool queueing, especially for long contexts and hybrid-model recurrent state. Replay extensions that associate phase placement and state transfers with each operation could evaluate these designs through complete-task SLOs.

\section{Related Work}
\label{sec:related}

\noindent\textbf{Multi-agent frameworks and collaboration evaluation.} AutoGen, MetaGPT, ChatDev, and Magentic-One implement collaborative workflows~\cite{autogen,metagpt,chatdev,magenticone}; AgentBench and MultiAgentBench evaluate agent outcomes and coordination~\cite{agentbench,multiagentbench}. These works motivate representative collaboration patterns, whereas MAServeBench examines how altering those patterns changes serving demand and task performance.

\noindent\textbf{Agentic serving and hardware benchmarking.} XPerf profiles observed agentic traces, reconstructs execution graphs, and synthesizes serving workloads by varying graph arrivals for DAG-aware, token-wise replay~\cite{xperf}. AgentReplay studies token-wise fidelity and optional router replay~\cite{agentreplay}. AgentSysBench characterizes complete applications while varying orchestration and deployment factors~\cite{agentsysbench}, while AgentPerfBench profiles agentic inference and kernel costs~\cite{agentperfbench}. Maestro, TokenCake, and TokenDance exploit workflow or context structure for serving~\cite{maestro,tokencake,tokendance}; Chakra standardizes execution traces~\cite{chakra}. MAServeBench instead makes collaboration semantics a workload-generation variable: shared task bindings reveal changes in calls, contexts, and typed dependencies, while hierarchical Semantic Traces preserve each realization for release and resource controls linked to serving, hardware, and task-SLO measurements.
\section{Conclusion}
MAServeBench provides a source-grounded, extensible benchmark for multi-agent serving. Composable stages vary collaboration and context construction under shared task bindings, generating distinct workloads through native execution. Hierarchical Semantic Traces preserve realized operations and dependencies for controlled comparisons of legal release and serving resources at fixed work. Task-linked measurements reveal how collaboration shapes parallelism, locality affects compute efficiency, and KV pressure impacts task completion. Decoupling workflow semantics from backend execution lets Semantic Traces serve as workloads for serving backends and architecture simulators, supporting MAS serving and hardware evaluation.

\section*{Acknowledgments}
We would like to express our sincere gratitude to Jianming Tong for his insightful discussions, thoughtful suggestions, and valuable feedback throughout the development of this work. His perspectives and constructive comments helped
us refine our understanding of the research challenges and improve the presentation of this study. We greatly appreciate his generous support
and encouragement.

\bibliography{example_paper}
\bibliographystyle{mlsys2025}

% \section*{AI Usage Disclosure}
% OpenAI ChatGPT assisted with parts of the experimental design and experimental code development. It also assisted with manuscript organization, English-language revision, literature-source checks, analysis of recorded measurements, figure preparation, and document compilation and consistency checks. The authors retain responsibility for the experimental methodology, reported results, citations, and final manuscript.

% The preprint includes the supplementary material after the references.
\clearpage
\appendix
\counterwithin{figure}{section}
\counterwithin{table}{section}
\counterwithin{equation}{section}
% Appendix-only float layout: pack evidence without flushing each section.
\raggedbottom
\setcounter{topnumber}{8}
\setcounter{bottomnumber}{4}
\setcounter{totalnumber}{12}
\setcounter{dbltopnumber}{4}
\renewcommand{\topfraction}{.98}
\renewcommand{\bottomfraction}{.95}
\renewcommand{\textfraction}{.02}
\renewcommand{\floatpagefraction}{.75}
\renewcommand{\dbltopfraction}{.98}
\renewcommand{\dblfloatpagefraction}{.75}
\setlength{\floatsep}{8pt plus 2pt minus 2pt}
\setlength{\textfloatsep}{10pt plus 2pt minus 2pt}
\setlength{\intextsep}{8pt plus 2pt minus 2pt}
\setlength{\dblfloatsep}{8pt plus 2pt minus 2pt}
\setlength{\dbltextfloatsep}{10pt plus 2pt minus 2pt}
\makeatletter
\setlength{\@fptop}{0pt}
\setlength{\@fpsep}{10pt}
\setlength{\@fpbot}{0pt plus 1fil}
\setlength{\@dblfptop}{0pt}
\setlength{\@dblfpsep}{10pt}
\setlength{\@dblfpbot}{0pt plus 1fil}
\makeatother

\section{Source Workflows and Functional Validation}
\label{app:sources}
This appendix supports \S\S\,\ref{sec:workload} and~\ref{sec:validity-workflows}: structural source mappings establish collaboration coverage, and paired native results connect source functionality to realized model and tool work.

\subsection{Documented Workflow Mappings}
\label{app:source-mappings}
The twelve inspected workflows cover four local templates and workflow composition. Table~\ref{tab:workflow-grounding} in \S\,\ref{sec:validity-workflows} gives the complete source-to-participant mappings.

The suite's demand population contains 210 native realizations, comprising 30 task bindings across seven structures. Figure~\ref{fig:app-natural} reports their input/output work. A separate fixed bank contains three bindings per structure and records operation counts, graph depth, level width, and typed edges. These records connect each executable instance to its template composition and input/output contract.

\begin{table}[H]
\centering\scriptsize
\setlength{\tabcolsep}{4pt}
\caption{Suite graph geometry. Each row summarizes three fixed-bank bindings with the same structure. Width counts model operations at a graph level. C1/C2 follow Figure~\ref{fig:eval-execution}.}
\label{tab:suite-geometry}
\begin{tabular}{@{}lrrrr@{}}
\toprule
Structure & Operations & Model calls & Depth & Width \\
\midrule
Spawn & 6 & 5 & 3 & 4 \\
Fork--join & 5 & 5 & 2 & 4 \\
Loop & 3 & 2 & 3 & 1 \\
Debate & 13 & 12 & 4 & 4 \\
C1 & 13 & 11 & 7 & 4 \\
C2 & 17 & 15 & 8 & 4 \\
Sequence & 16 & 16 & 6 & 3 \\
\bottomrule
\end{tabular}
\end{table}

\subsection{Reference Workflows and Paired Functional Validation}
\label{app:functional-validation}
\noindent\textbf{Frameworks and task sets.}
Six reference frameworks form seven paired comparisons across four task sets: Agyn and SWE-Debate on SWE-bench Verified, Magentic-One and OWL Workforce on GAIA, Magentic-One on WebArena, and GPT Researcher and Open Deep Research on DeepResearch Bench~\cite{agyn,swe_debate,magenticone,owl_workforce,gpt_researcher_multiagent,open_deep_research}. Together, these pairs cover worker orchestration, parallel research, review and revision, and structured debate.

\noindent\textbf{Paired evaluation.}
Each reference--realization pair uses the same task instances, Qwen3.5-9B, and evaluation procedure, with comparable tool access and execution budgets. Both arms execute natively: generated answers and tool results determine subsequent work. Quality is measured by resolved rate on SWE-bench Verified, answer accuracy on GAIA, task success on WebArena, and RACE report quality on DeepResearch Bench. Table~\ref{tab:app-functional-validity} reports quality together with mean model calls, tool calls, and cumulative input/output tokens per attempted task, aggregated across participants.

\noindent\textbf{Workflow correspondence.}
The realizations preserve source-relevant control, context delivery, and tool dependencies. Worker orchestration maps to Spawn; independent research and aggregation map to Fork--Join; review and revision map to Refinement Loop. SWE-Debate combines candidate localization, structured Debate, and iterative patch search. Workflow composition connects these operations into the corresponding task procedure.

\begin{table*}[!t]
\centering
\scriptsize
\setlength{\tabcolsep}{3.8pt}
\renewcommand{\arraystretch}{1.13}
\caption{Paired functional validation of source MAS workflows and their MAServeBench realizations using \textbf{Qwen3.5-9B}. Each pair shares $N$ task instances, model settings, and evaluation procedures. Quality measures resolved rate (SWE-bench Verified), answer accuracy (GAIA), task success (WebArena), or RACE report quality (DeepResearch Bench). LLM/tool calls and input/output tokens (in thousands) are averaged per attempted task. Quality scores across datasets are not directly comparable.}
\label{tab:app-functional-validity}

\begin{tabularx}{\textwidth}{@{}
>{\raggedright\arraybackslash}p{.13\textwidth}
>{\raggedright\arraybackslash}p{.115\textwidth}
>{\raggedright\arraybackslash}X
>{\centering\arraybackslash}p{.08\textwidth}
>{\centering\arraybackslash}p{.06\textwidth}
>{\centering\arraybackslash}p{.06\textwidth}
>{\centering\arraybackslash}p{.08\textwidth}
>{\centering\arraybackslash}p{.08\textwidth}@{}}

\toprule
\textbf{Framework} &
\textbf{Execution} &
\textbf{Workflow / abstraction} &
\textbf{Quality $\uparrow$} &
\textbf{LLM} &
\textbf{Tools} &
\textbf{Input (k)} &
\textbf{Output (k)} \\
\midrule

\multicolumn{8}{@{}l}{\textbf{SWE-bench Verified}\quad $N=500$\quad Resolved rate (\%)} \\
\addlinespace[2pt]

\textbf{Agyn} & Reference & Original issue-resolution team & 60.4\% & 37 & 125 & 1,156 & 175 \\
\rowcolor{gray!10}
& MAServeBench & Spawn + Refinement Loop & 62.4\% & 41 & 129 & 1,074 & 168 \\

\addlinespace[3pt]

\textbf{SWE-Debate} & Reference & Localization, debate, and MCTS patch search & 55.6\% & 52 & 146 & 1,066 & 167 \\
\rowcolor{gray!10}
& MAServeBench & Candidate generation + Debate + iterative search & 53.4\% & 48 & 143 & 989 & 174 \\

\midrule
\multicolumn{8}{@{}l}{\textbf{GAIA (dev)}\quad $N=165$\quad Answer accuracy (\%)} \\
\addlinespace[2pt]

\textbf{Magentic-One} & Reference & Orchestrator and selected workers & 62.4\% & 37 & 72 & 376 & 64 \\
\rowcolor{gray!10}
& MAServeBench & Spawn + workflow composition & 61.2\% & 35 & 74 & 356 & 48 \\

\addlinespace[3pt]

\textbf{OWL Workforce} & Reference & Planner, coordinator, and workers & 64.8\% & 67 & 64 & 1,236 & 89 \\
\rowcolor{gray!10}
& MAServeBench & Spawn + workflow composition & 67.9\% & 66 & 64 & 1,098 & 85 \\

\midrule
\multicolumn{8}{@{}l}{\textbf{WebArena}\quad $N=812$\quad Task success (\%)} \\
\addlinespace[2pt]

\textbf{Magentic-One} & Reference & Orchestrator and selected workers & 44.7\% & 56 & 97 & 387 & 89 \\
\rowcolor{gray!10}
& MAServeBench & Spawn + workflow composition & 40.9\% & 60 & 104 & 402 & 97 \\

\midrule
\multicolumn{8}{@{}l}{\textbf{DeepResearch Bench}\quad $N=100$\quad RACE report quality (fixed evaluator)} \\
\addlinespace[2pt]

\textbf{GPT Researcher} & Reference & Parallel research and section review & 0.498 & 99 & 147 & 1,344 & 97 \\
\rowcolor{gray!10}
& MAServeBench & Fork--Join + Refinement Loop & 0.479 & 105 & 153 & 1,367 & 94 \\

\addlinespace[3pt]

\textbf{Open Deep Research} & Reference & Supervisor and parallel researchers & 0.4509 & 15 & 47 & 446 & 53 \\
\rowcolor{gray!10}
& MAServeBench & Spawn + Fork--Join & 0.4608 & 17 & 47 & 441 & 56 \\

\bottomrule
\end{tabularx}
\end{table*}

\noindent\textbf{Results.}
The paired realizations track their source workflows in task quality: absolute differences are 1.2--3.8 percentage points on the success/accuracy measures and at most 0.019 on RACE. Execution demand varies within this quality range. For example, GPT Researcher's realization invokes more tools, while OWL's realization uses fewer input tokens with nearly the same call counts. Across the pairs, task functionality is accompanied by distinct model, tool, and context-processing demands. These measurements ground the realizations used to study collaboration as a serving workload.

\begin{figure}[!t]
\centering
\includegraphics[width=\columnwidth]{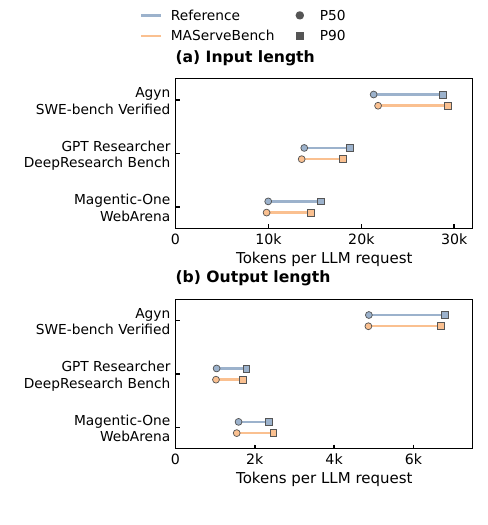}
\vspace{-20pt}
\caption{\textbf{Request lengths in source workflows and MAServeBench realizations.} Reported (a) input and (b) output tokens per LLM request for the paired native evaluations in Table~\ref{tab:app-functional-validity}. Circles and squares denote P50 and P90 over requests; lines join the two distribution percentiles. Colors distinguish reference and realization.}
\label{fig:source-serving-demand}
\end{figure}

\noindent\textbf{Request-length correspondence.}
Figure~\ref{fig:source-serving-demand} complements per-task totals with author-reported input/output P50/P90 over the requests in three paired native evaluations from Table~\ref{tab:app-functional-validity}, using its model and token-accounting protocol.

Input medians differ by at most 2.3\% across the three pairs; the largest input P90 change is a 7.1\% reduction for Magentic-One on WebArena. Output P90 is lower in the Agyn and GPT Researcher realizations but 5.0\% higher for Magentic-One. The realizations retain similar input medians with workflow-dependent tail shifts.

\subsection{Source-to-Serving Workload Correspondence}
\label{app:source-serving-correspondence}

\begin{figure}[!t]
\centering
\includegraphics[width=\columnwidth]{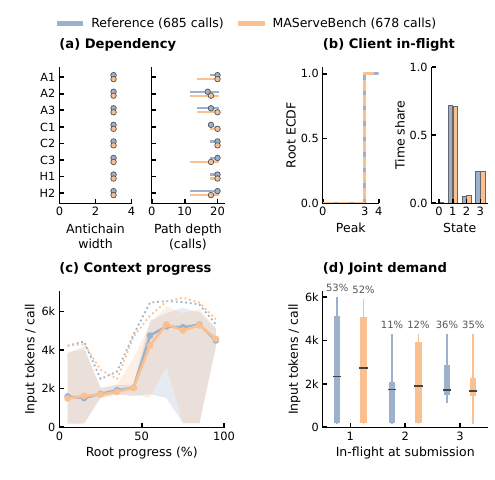}
\vspace{-20pt}
\caption{\textbf{Source-to-MAServeBench serving-workload correspondence.} Eight bindings, three blocks, 24 roots per arm. (a) Antichain width and path depth: binding medians and block ranges. (b) Root-local peak client in-flight ECDF and mean per-root time shares, distinct from batch size. (c) Input tokens over ten progress bins: weighted median, P25--P75 band, dotted P90. (d) Tokens by in-flight at dispatch: median, P25--P75 thick and P10--P90 thin bars; percentages are root-weighted request shares. Empty bins remain unconnected; weighting is defined below.}
\label{fig:source-serving-correspondence}
\end{figure}

\begin{table*}[!t]
\centering\scriptsize
\setlength{\tabcolsep}{5pt}
\caption{Paired serving-workload statistics. Each arm reports median [P25, P75] across eight binding-level medians over three blocks. Paired differences first take the median of three within-binding block differences, then summarize the eight bindings. The last column gives arm-wise P90 over binding medians. Retrieval calls use the shared frozen-document tool boundary; graph metrics use LLM calls. Session rows first summarize eligible sessions within each root; changes and ratios are formed per session before aggregation.}
\label{tab:source-serving-correspondence}
\begin{tabular*}{\textwidth}{@{\extracolsep{\fill}}lrrrr@{}}
\toprule
Metric & Reference & MAServeBench & Paired difference & P90 (R/M) \\
\midrule
Model calls & 30 [28, 30] & 30 [28, 30] & 0 [0, 0.2] & 30/30 \\
Retrieval calls & 4 [4, 4] & 4 [4, 4] & 0 [0, 0] & 4/4 \\
LLM required edges & 31 [29, 31] & 31 [29, 31] & 0 [0, 0.2] & 31/31 \\
LLM depth (calls) & 20 [18, 20] & 20 [18, 20] & 0 [0, 0.2] & 20/20 \\
LLM level width & 3 [3, 3] & 3 [3, 3] & 0 [0, 0] & 3/3 \\
LLM antichain width & 3 [3, 3] & 3 [3, 3] & 0 [0, 0] & 3/3 \\
Peak client in-flight & 3 [3, 3] & 3 [3, 3] & 0 [0, 0] & 3/3 \\
Total input (k tokens) & 75.2 [73.3, 77.0] & 72.7 [70.3, 75.6] & -1.8 [-4.5, 3.0] & 82.2/79.8 \\
Total output (k tokens) & 20.8 [19.7, 22.3] & 20.6 [19.2, 21.0] & 0.2 [-1.3, 1.6] & 24.0/22.5 \\
Request input P50 (tokens) & 1904.5 [1824.9, 1998.1] & 1777.2 [1744.9, 1944.8] & -54 [-193.6, 13.5] & 2108.4/2163 \\
Request input P90 (tokens) & 5181.5 [4989.5, 5800.5] & 5235.3 [4990.3, 5417.1] & -11.7 [-176.4, 81.1] & 5904.6/5833.2 \\
Root latency (s) & 306.6 [289.4, 325.9] & 304.6 [279.2, 315.2] & 2.4 [-3.5, 30.9] & 355.9/342.8 \\
Multi-call sessions & 10 [9.8, 10] & 10 [9, 10] & 0 [-0.2, 0.2] & 10/10 \\
Session first input (tokens) & 1468.5 [1363.8, 1575.5] & 1410.8 [1331.2, 1511.8] & -38.2 [-103.9, -25] & 1623.8/1575.8 \\
Session last input (tokens) & 3139 [2927.5, 3500.2] & 3072.2 [2972, 3253.4] & 107.8 [-170, 240.5] & 3600.9/3462.5 \\
Session input change (tokens) & 258.5 [158, 311.9] & 265 [212, 318.1] & 60.5 [30.5, 110.9] & 337.3/436.4 \\
Session last/first input & 1.11 [1.09, 1.13] & 1.14 [1.12, 1.17] & 0.02 [0, 0.08] & 1.14/1.27 \\
\bottomrule
\end{tabular*}
\end{table*}

\noindent\textbf{Source and shared inputs.}
We execute GPT Researcher's original ChiefEditor and Editor LangGraph orchestration at revision \texttt{6f998577\allowbreak{}d547b1e5\allowbreak{}4ec662da\allowbreak{}c63583aa\allowbreak{}11e3b84b}~\cite{gpt_researcher_multiagent}, retaining planning, parallel section research, draft refinement, writing/checking, and publication. Both implementations use its original role prompts and the same frozen-document retrieval adapter. The MAServeBench realization composes Fork--Join with Refinement Loop and selects the reviewed draft fields, matching the source Editor's aggregation interface.
Eight predeclared questions use the existing official-document snapshots: cancellation, scheduling, and shutdown for asyncio (A1--A3); CPU control, memory control, and delegation for cgroup v2 (C1--C3); and freshness and validation for HTTP caching (H1--H2). This comparison studies the pinned workflow's local-document profile across these eight bindings and three document domains.

\noindent\textbf{Paired execution and observation.}
Reference and MAServeBench share Qwen3.5-9B, its tokenizer and chat template, one RTX PRO 5000, vLLM 0.30.0, BF16, TP1, a 64-request limit, temperature zero, a 2,048-token output budget, and disabled prefix caching. Three blocks use shared seeds 0/1/2 and alternating Reference/MAServeBench order AB/BA/AB. Each block contains two four-root cohorts with identical relative arrivals at 0/0.1/0.2/0.3 seconds. Queues and live KV state drain before each arm. Both workflows natively generate their own calls and outputs.
The paired experiment completes 48 roots, with 685 Reference and 678 MAServeBench model calls. A single-host monotonic clock records dispatch and streaming completion; root latency spans scheduled arrival to native runner return. Backend input-token counts are independently checked by applying the same chat template to each actual transmitted prompt. Stage and session identities follow the original stage/role interfaces; framework participant IDs and complete artifacts remain in the records.

\noindent\textbf{Metric definitions.}
Required edges follow actual source sequencing, fork/join and feedback routes, or recorded MAServeBench dependencies. We retain LLM calls, collapse intervening non-LLM paths, and remove transitive edges. Depth counts calls on the longest path; level width uses longest-path topological levels; antichain width uses reachability and bipartite matching. Client in-flight counts submitted, unfinished requests and includes the current request at dispatch. Within-session change compares the first and last prompts only in sessions with at least two calls.
There are 230 of 336 eligible Reference sessions and 229 of 336 MAServeBench sessions. For progress-bin token distributions, each binding has equal weight, divided equally among its observed roots and their calls in that bin. Joint concurrency--context distributions give each root equal total weight. Block repeats remain paired within a binding.

\noindent\textbf{Observed correspondence.} Both arms realize antichain width 3, depth 12--20 calls, and root-local in-flight peak 3 (Figure~\ref{fig:source-serving-correspondence}, Table~\ref{tab:source-serving-correspondence}). Three requests overlap for 23.0\%/23.0\% of root lifetime on average. Both arms show a common increase in median prompt length toward later workflow progress, with local variation in context timing.

\FloatBarrier
\section{Trace and Replay Conformance}
\label{app:execution-validation}
This appendix supports \S\S\,\ref{sec:trace-execution} and~\ref{sec:validity-semantics} through backend-token, typed-dependency, and replay checks.

\subsection{Request and Dependency Checks}
The execution audit links each completed model operation to backend token counts and checks the release of dependent operations against predecessor completion. Across the model study, all 1,032 completed requests match their recorded input/output counts. All 186 recorded predecessor edges satisfy the release constraint with a 1\,ms timestamp tolerance. Attempt-level records retain completion status and consumed work.

\noindent\textbf{Bounded replay calibration.}
The paired calibration starts from a completed native execution. Its replay fixes the resolved request payloads, realized output lengths, tool results, and predecessor edges, then releases each request when its recorded dependencies complete. Native and replay arms share the warmed serving configuration. The comparison records complete-task latency alongside request-level token and dependency checks. This separates replay conformance from the functional comparisons in Appendix~\ref{app:functional-validation}.

% EVALUATION_SUPPORT_20261021
On one 910C logical device, five isolated native traces and ten paired replays use prefix caching off and the same non-thinking chat template. The audit preserves 48 typed edges and verifies 46 replay requests (Table~\ref{tab:replay-calibration}); the earlier thinking-enabled f\&j trace remains separately listed.

\begin{table}[!t]
\centering\small
\caption{Bounded native-to-replay calibration. Error is absolute task-latency difference divided by native task latency. Each native trace has two replays.}
\label{tab:replay-calibration}
\begin{tabular}{@{}lrrrr@{}}
\toprule
Structure & Native & Replay & Median & Maximum \\
\midrule
Composed & 2 & 4 & 0.118\% & 0.172\% \\
f\&j & 1 & 2 & 0.136\% & 0.145\% \\
Loop & 2 & 4 & 0.471\% & 0.797\% \\
Earlier f\&j & 1 & 2 & 0.577\% & 0.658\% \\
\bottomrule
\end{tabular}
\end{table}

\begin{table}[!t]
\centering\scriptsize
\setlength{\tabcolsep}{3pt}
\caption{Replay-calibration attempt ledger. The earlier protocol includes one failed native refinement-loop attempt; successful traces retain their paired replays.}
\label{tab:replay-attempts}
\begin{tabular}{lrrr}
\toprule
Protocol & Attempts & Eligible & Replays \\
\midrule
Earlier thinking on & 2 & 1 & 2 \\
Continuation thinking off & 5 & 5 & 10 \\
Total & 7 & 6 & 12 \\
\bottomrule
\end{tabular}
\end{table}

\subsection{Concurrent and Cache-Enabled Replay}
\label{app:replay-load}
The Qwen3-8B/910C extension pairs native and length-locked replay roots for fork--join and fork--join with refinement, with one or 16 roots per cohort and prefix caching disabled or enabled. Native and replay arms start from confirmed cold cache. Each arm preserves canonical model inputs, actual input/output token counts, operation identities, and typed predecessors. All 204 root pairs pass these work checks; required predecessors have no early releases. Table~\ref{tab:replay-load} reports timing separately as replay/native minus one. Each stratum contains three blocks.

\begin{table}[!t]
\centering\scriptsize
\setlength{\tabcolsep}{3pt}
\caption{Replay under concurrency and prefix caching. Median error summarizes paired block means; the range retains individual root errors. FJ: fork--join; FJR: fork--join with refinement.}
\label{tab:replay-load}
\begin{tabular}{@{}lrlrrr@{}}
\toprule
Case & Roots & Cache & Pairs & Median & Root range \\
\midrule
FJ & 1 & Off & 3 & $-0.33\%$ & $[-0.58,-0.01]\%$ \\
FJ & 1 & On & 3 & $-0.80\%$ & $[-0.93,+0.01]\%$ \\
FJ & 16 & Off & 48 & $+0.20\%$ & $[-1.74,+2.88]\%$ \\
FJ & 16 & On & 48 & $-0.38\%$ & $[-1.84,+0.74]\%$ \\
FJR & 1 & Off & 3 & $-0.64\%$ & $[-0.66,-0.17]\%$ \\
FJR & 1 & On & 3 & $-0.50\%$ & $[-1.13,-0.37]\%$ \\
FJR & 16 & Off & 48 & $-0.25\%$ & $[-1.85,+0.35]\%$ \\
FJR & 16 & On & 48 & $-0.29\%$ & $[-1.58,+0.76]\%$ \\
\bottomrule
\end{tabular}
\end{table}

\subsection{Optional Exact-Token Replay}
\label{app:token-replay}
The optional adapter calls vLLM 0.30.0's official trace replay interface on CUDA Model Runner V2, without modifying the serving package. A finite Qwen3-4B-Instruct-2507/RTX PRO 5000 gate records 16 native roots (66 model requests) across Fork--Join and Fork--Join with Refinement using the preceding artifact, one or three concurrent roots, and prefix caching off/on. It then executes 198 token-locked and 198 length-locked requests in three paired blocks per setting, resetting prefix cache before each arm. All 16,077 recorded output IDs match position by position; engine output updates match the generated-token counts, and an independent fixture containing EOS verifies termination handling. Inputs, operation identities, recorded artifacts, and typed predecessors are retained.

\begin{table}[!t]
\centering\scriptsize
\caption{Optional token/length replay wall-time ratios on the same engine and runner. Each cell is a three-block median [minimum, maximum]. FJR consumes the preceding Fork--Join artifact.}
\label{tab:token-replay-overhead}
\begin{tabular}{@{}llrr@{}}
\toprule
Workflow & Roots & Cache off & Cache on \\
\midrule
FJ & 1 & 0.992 [0.987, 1.003] & 0.995 [0.986, 0.997] \\
FJ & 3 & 0.994 [0.993, 1.003] & 0.994 [0.989, 0.997] \\
FJR & 1 & 1.000 [0.991, 1.002] & 1.000 [0.993, 1.001] \\
FJR & 3 & 0.993 [0.991, 0.995] & 0.999 [0.990, 1.004] \\
\bottomrule
\end{tabular}
\end{table}

The evaluated CUDA Model Runner V2 adapter supports token-locked replay; the current HTTP and Ascend adapters use length-locked replay. Preflight rejects absent or re-tokenized output IDs, model/tokenizer/template mismatches, invalid IDs, context overflow, and unsupported decoding features. The verified gate covers the stated workflows, cache settings, concurrency, and EOS handling; forced KV preemption and live cancellation require separate validation. The ratios summarize the tested CUDA configuration and workload cells across all paired blocks.

\section{Systems, Profiling, and Metric Contracts}
\label{app:measurement}
This appendix supports \S\,\ref{sec:eval-setup} and the measurement units used in \S\S\,\ref{sec:eval-control}--\ref{sec:eval-slo}. Tables~\ref{tab:protocol-map} and~\ref{tab:counter-contract} link the model/platform scopes to their measurement contracts.
We retain the completed physical windows and their original request, operation, and profiling records. Unprofiled application windows provide latency; native captures provide the costs within their recorded intervals. A pure-decoding step schedules one token per request with each request already at the end of its prompt. Batch distributions weight each step by its duration, stop at the last model-request completion, and normalize within a run before combining blocks. The final-drain boundary is the first scheduling time of the cohort's final new model request. Native operator shares sum kernel durations by family within a capture.

GPU event timings and counter measurements come from independent passes; RTX PRO 5000 timing and counter passes use different physical devices of the same SKU. NPU architecture comparisons match each marked operator to the corresponding native Memory pass on both devices, with eight calls per case and pass. Pipeline and cache passes are retained as supplementary observations. The 950DT application captures retain their shared-environment classification; cross-generation performance comparisons use the matched operator protocol.

\begin{table*}[!t]
\centering\scriptsize
\setlength{\tabcolsep}{3pt}
\caption{Application and operator protocol map. Application rows give concurrent root users and request limits; operator rows give execution shape. Timing windows and native captures are retained separately.}
\label{tab:protocol-map}
\begin{tabular}{llll}
\toprule
Study & Model / platform & Users & Limit/shape \\
\midrule
Demand (Fig.~\ref{fig:eval-demand}) & Qwen3-8B / 910C & 32/64/128 & 128 \\
Execution shapes (Fig.~\ref{fig:eval-execution}) & Qwen3-8B / 910C & 64 & 128 \\
Attention captures (Fig.~\ref{fig:eval-npu-memory}a) & Qwen3-8B / 910C & 64 & 64 \\
State pressure (Fig.~\ref{fig:eval-state}) & Qwen3-8B / 910C & 64 & 128 \\
Stage/root (Fig.~\ref{fig:eval-stages}) & Qwen3.5-9B / RTX PRO 5000 & 64 & 64 \\
NPU hybrid captures (Sec.~\ref{sec:eval-hardware}) & Qwen3.5-9B / 910C and 950DT & 32 & 32 \\
GPU projections (Figs.~\ref{fig:eval-gpu-compute}) & Qwen3.5-9B shapes / A800 and RTX PRO 5000 & operator & rows 1/8; 512 control \\
NPU compute (Fig.~\ref{fig:eval-npu-compute}) & Matrix/auxiliary probes / 910C and 950DT & operator & matched shapes/paths \\
NPU attention (Fig.~\ref{fig:eval-npu-memory}b) & Qwen3-8B geometry / 910C and 950DT & operator & batch 32 \\
Optional control (Fig.~\ref{fig:eval-release}) & Qwen3-8B / 910C & 12 & four calls/root \\
Fixed targets (Fig.~\ref{fig:eval-cotenancy}) & Qwen3-8B / 910C & 9+55 & 128 \\
Unmatched targets (Fig.~\ref{fig:app-target-unmatched}) & Qwen3-8B / 910C & 9+55 & 128 \\
KV budgets (Fig.~\ref{fig:eval-budget}) & Qwen3-8B / 910C & 64 & 128; 16/32/40 GiB \\
Release/KV SLOs (Fig.~\ref{fig:eval-slo}) & Qwen3-8B / 910C & external arrivals & 128; 16/40 GiB \\
Request-limit sensitivity (Fig.~\ref{fig:app-request-limit}) & Qwen3-8B / 910C & external arrivals & 64/128 \\
\bottomrule
\end{tabular}
\end{table*}

\begin{table*}[!t]
\centering\scriptsize
\setlength{\tabcolsep}{3pt}
\caption{Measurement contracts. Numerators and denominators refer to the same stated observation scope.}
\label{tab:counter-contract}
\begin{tabular}{lp{.25\textwidth}p{.43\textwidth}}
\toprule
Quantity & Scope and unit & Statistic \\
\midrule
Arithmetic intensity & Qwen3-8B scheduler steps; FLOP/byte & $F/D$ for scheduled work and logical traffic; progress-bin duration weighting; aggregate $\sum F/\sum D$ \\
Allocated KV & Scheduler pool; GiB & Allocated blocks times block capacity; residency integrates occupancy over cohort wall time \\
Kernel-duration share & Native capture; percent & Duration sum of one mutually exclusive kernel family divided by all classified kernel duration in that capture \\
Projection time & Matched primitive; microseconds & CUDA events on GPUs; corresponding native Memory-pass ranges on NPUs \\
Physical DRAM reads & Counter pass; bytes per complete primitive & All constituent kernels summed per marked primitive, then median across complete samples; one BF16 weight normalizes the plotted reads \\
Stage and root latency & Unprofiled run; seconds & Within-binding means balanced across common task bindings; paired block ratios summarized by their median \\
Completion-path accounts & Observed dependency path; seconds & Latest-finishing predecessor path, with ready-to-submit, submit-to-first-schedule, scheduled residence, and boundary intervals \\
\bottomrule
\end{tabular}
\end{table*}

\subsection{Arrival-rate and Task-SLO Contracts}
For arrival-rate studies, the benchmark interface defines task goodput and SLO-constrained capacity over tested loads that pass completion and backlog guards.
\begin{equation}
\begin{aligned}
G_{\mathrm{task}}
&= \frac{N_{\mathrm{SLO\text{-}compliant\ tasks}}}{T}, \\
\lambda^{*}
&= \max\{\lambda\in\Lambda_{\mathrm{adm}} \mid A(\lambda)\geq\alpha\},
\end{aligned}
\label{eq:benchmark_metrics}
\end{equation}
where $T$ is the measurement interval, $A(\lambda)$ is the minimum attempted-task SLO attainment across required blocks and active task classes at load $\lambda$, and $\alpha$ is the required target. $\Lambda_{\mathrm{adm}}$ contains tested loads passing completion and backlog guards. Failed and incomplete tasks count as attempts. \emph{Resource efficiency} is SLO-compliant task goodput per accelerator.

\subsection{Legal Release, KV Capacity, and Shared Deadlines}
\label{app:release-kv}
\label{app:deadline-sensitivity}
This study crosses DAG/Barrier release with nominal 16/40\,GiB KV pools on Qwen3-8B/910C. The actual allocations are 15.996094 and 39.990234\,GiB. BF16, tensor parallelism one, a single serving instance, disabled prefix caching, and the 8192-token model and batched-token limits match the fixed-target study. Server and client request limits remain 128 in every arm. Each window starts after common warmup with running requests, waiting requests, and KV usage cleared.

\noindent\textbf{Stage-local legal control and fixed work.} Each W-DFR root executes 13 operations, including 11 LLM calls. In its fork--join stage, four independent LLM analyses read the same immutable context, with no peer-artifact reads or mutable-state operations; the reducer requires all four original branches. DAG retains the recorded graph. Barrier adds three sequential control edges among these analyses, carrying the stage identity and coordination annotation. The intervention preserves every original dependency and passes checks for independence, acyclicity, and typed data/artifact/state relationships. Canonical requests, output-length targets, root IDs, bindings, and arrival times match within each four-arm comparison. Recorded downstream inputs remain fixed; generated outputs are retained in the execution records.

\noindent\textbf{Arrivals and a common locked deadline.} Nine independent low-load, single-root DAG/40\,GiB reference windows have root P95 241.134650 seconds, giving a shared $1.5\times$ deadline of 361.701975 seconds. Reference and formal root P95 use linear interpolation at zero-based sorted index $.95(n-1)$. The formal grid uses four Poisson rates, 0.00677316, 0.01625559, 0.02709266, and 0.04334825 roots/s, with arrival seeds 246600, 246601, and 246602 for its three paired blocks. All cells use $T=1181.132588$ seconds, fixed by $T=\max(600,8/\lambda_{\min})$. Arrivals and physical device match across all four arms of each rate/block. The 48 formal windows contain 1,492 root executions and 16,412 LLM calls. Including the nine reference, six fixed-target, and eight pilot windows gives 71 completed windows, 1,981 root executions, and 18,353 model calls, with no failures or incomplete roots. Reference and pilot windows are excluded from the formal three-block summaries. The roots repeatedly instantiate the three frozen bindings listed above.

\noindent\textbf{Counting complete tasks.} Arrival-window goodput counts roots arriving in $[0,T)$ that eventually meet their individual deadline, divided by $T$; a timely root may finish during drain after $T$. Attainment divides the same timely-root count by all roots arriving within the window. Output TPS and drain-inclusive timely-root rate instead use $\max(T,T_{\mathrm{cohort}})$. Finite Poisson cohorts can contain more roots than $\lambda T$, so an observed arrival-window goodput may exceed nominal offered rate. Table~\ref{tab:release-kv-deadlines} preserves the three-block medians and ranges.

\begin{table*}[!t]
\centering\scriptsize
\setlength{\tabcolsep}{3pt}
\caption{Legal-release/KV deadline outcomes: median/minimum/maximum across three paired blocks. Goodput is timely roots per arrival-window second; attainment is the fraction meeting the shared deadline.}
\label{tab:release-kv-deadlines}
\begin{tabular}{@{}llrll@{}}
\toprule
Rate (roots/s) & Release & KV (GiB) & Goodput: med/min/max & Attainment: med/min/max \\
\midrule
0.00677316 & DAG & 16 & 0.008466/0.005927/0.010160 & 1.000/1.000/1.000 \\
0.00677316 & DAG & 40 & 0.008466/0.005927/0.010160 & 1.000/1.000/1.000 \\
0.00677316 & Barrier & 16 & 0.002540/0.002540/0.005080 & 0.429/0.300/0.500 \\
0.00677316 & Barrier & 40 & 0.003387/0.002540/0.003387 & 0.400/0.333/0.429 \\
0.01625559 & DAG & 16 & 0.018626/0.014393/0.021166 & 1.000/0.917/1.000 \\
0.01625559 & DAG & 40 & 0.018626/0.014393/0.021166 & 1.000/0.917/1.000 \\
0.01625559 & Barrier & 16 & 0.000847/0.000000/0.000847 & 0.040/0.000/0.059 \\
0.01625559 & Barrier & 40 & 0.000847/0.000000/0.000847 & 0.040/0.000/0.059 \\
0.02709266 & DAG & 16 & 0.004233/0.002540/0.020319 & 0.122/0.068/0.923 \\
0.02709266 & DAG & 40 & 0.003387/0.002540/0.019473 & 0.098/0.068/0.885 \\
0.02709266 & Barrier & 16 & 0.000000/0.000000/0.000000 & 0.000/0.000/0.000 \\
0.02709266 & Barrier & 40 & 0.000000/0.000000/0.000000 & 0.000/0.000/0.000 \\
0.04334825 & DAG & 16 & 0.000847/0.000000/0.001693 & 0.015/0.000/0.047 \\
0.04334825 & DAG & 40 & 0.000000/0.000000/0.001693 & 0.000/0.000/0.047 \\
0.04334825 & Barrier & 16 & 0.000000/0.000000/0.000000 & 0.000/0.000/0.000 \\
0.04334825 & Barrier & 40 & 0.000000/0.000000/0.000000 & 0.000/0.000/0.000 \\
\bottomrule
\end{tabular}
\end{table*}

\begin{figure*}[!t]
\centering
\includegraphics[width=\textwidth]{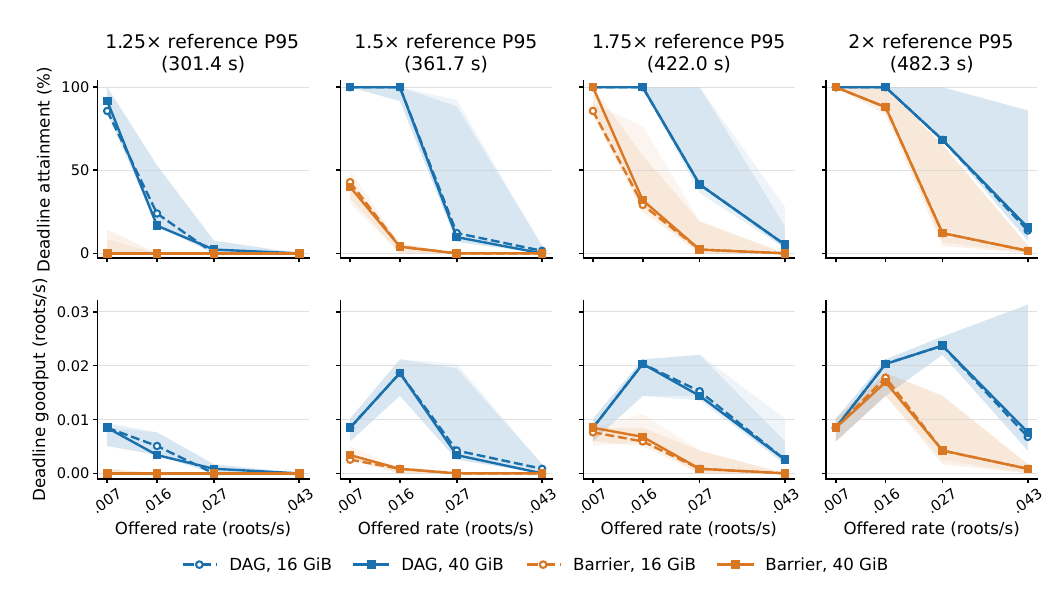}
\vspace{-20pt}
\caption{\textbf{Offline deadline sensitivity over the complete formal grid.} Columns use four multiples of the same independent DAG/40\,GiB reference P95. Top: task attainment; bottom: arrival-window deadline goodput. Lines/markers show three-block medians and shaded bands span minimum to maximum for each configuration. Exact offered rates and unchanged arrival/drain counting follow Table~\ref{tab:release-kv-deadlines}. Every task is re-scored from its original arrival/completion timestamps; no serving execution is repeated.}
\label{fig:app-deadline-sensitivity}
\end{figure*}

\noindent\textbf{Offline deadline sensitivity.} We retain the 48 execution windows, all 1,492 recorded root arrivals/completions, and the original counting interval, changing only the deadline multiplier. The same independent reference P95 gives deadlines 301.418313, 361.701975, 421.985638, and 482.269301 seconds at $1.25/1.5/1.75/2.0\times$. Figure~\ref{fig:app-deadline-sensitivity} includes every rate, release rule, KV budget, and paired block through median/range summaries. Re-scoring at $1.5\times$ reproduces all original timely counts and goodput values exactly.

At the second rate, DAG's median attainment is 24/16.7\% (16/40\,GiB) at $1.25\times$, and 100\% at all larger deadlines; Barrier gives 0\%, 4\%, 29.2/32\%, and 88\%, respectively. Under the main deadline, DAG's block range is 91.7--100\%, including the $22/24$ block, and Barrier's is 0--5.9\%. DAG has higher median attainment and goodput at this rate for all four deadlines and both budgets. Across all 96 paired block contrasts in the full sensitivity grid, DAG is never lower. Ties occur when both arms meet the loose deadline at the lowest rate or miss the strict deadline at the highest rate; at the second rate one block ties under $2.0\times$.

\begin{figure}[!t]
\centering
\includegraphics[width=\columnwidth]{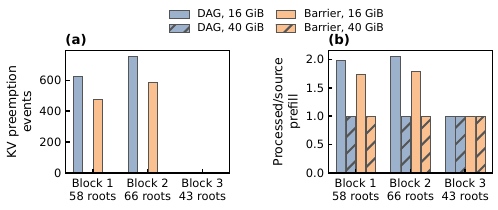}
\vspace{-20pt}
\caption{\textbf{High-load KV recomputation.} (a) Scheduler preemption events. (b) Processed/source prefill. Every configuration retains all three paired blocks, including the 43-root block without preemption.}
\label{fig:app-s66-recomputation}
\end{figure}

\begin{figure}[!t]
\centering
\includegraphics[width=\columnwidth]{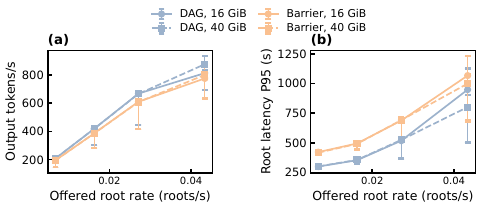}
\vspace{-20pt}
\caption{\textbf{Throughput and task tails over offered load.} (a) Output TPS over the drain-inclusive period. (b) Root P95 latency. Points show three-block medians; whiskers span their ranges.}
\label{fig:app-s66-tokens-tails}
\end{figure}

\begin{figure}[!t]
\centering
\includegraphics[width=\columnwidth]{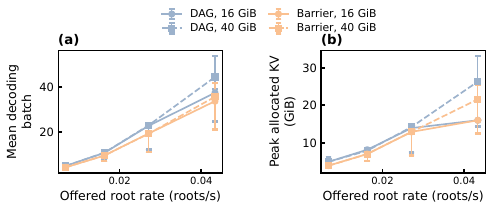}
\vspace{-20pt}
\caption{\textbf{Batch exposure and state over offered load.} (a) Decoding batch weighted by pure-decode step duration through final request completion. (b) Peak allocated KV occupancy. Points show three-block medians and ranges.}
\label{fig:app-s66-batch-state}
\end{figure}

\noindent\textbf{High-load state and service costs.} The highest-rate blocks contain 58, 66, and 43 roots, respectively, with identical arrivals across configurations. Table~\ref{tab:release-kv-high-load} and Figure~\ref{fig:app-s66-recomputation} retain the third block with no preemption alongside the two pressured blocks. Figures~\ref{fig:app-s66-tokens-tails} and~\ref{fig:app-s66-batch-state} show service and state over the full rate grid. The first three rates have no preemptions and a prefill factor of 1.0 in every configuration. At the highest rate, a larger pool eliminates preemption and repeated prefill, increasing the retained in-flight state and mean decoding batch.

\begin{table*}[!t]
\centering\scriptsize
\setlength{\tabcolsep}{3pt}
\caption{Highest-rate state and service outcomes. Preemptions retain blocks 1/2/3; other columns are three-block medians. These are marginal summaries, not paired-effect ratios.}
\label{tab:release-kv-high-load}
\begin{tabular}{@{}lrllrr@{}}
\toprule
Release & KV (GiB) & Preemptions & Prefill factor & Root P95 (s) & Output TPS \\
\midrule
DAG & 16 & 628/754/0 & 1.981 & 948.7 & 813.3 \\
DAG & 40 & 0/0/0 & 1.000 & 801.8 & 876.2 \\
Barrier & 16 & 480/590/0 & 1.748 & 1067.3 & 775.4 \\
Barrier & 40 & 0/0/0 & 1.000 & 999.6 & 800.5 \\
\bottomrule
\end{tabular}
\end{table*}

Within-block 40/16\,GiB ratios, followed by the median across all three blocks, give DAG root P95 $0.8451\times$, output TPS $1.0774\times$, and mean decode batch $1.1884\times$. Barrier gives $0.9366\times$, $1.0324\times$, and $1.0582\times$, respectively. These paired ratios differ from ratios of marginal medians. DAG's paired arrival-window goodput changes are $[0,-0.000846645,0]$ roots/s; Barrier's are all zero. The larger pool changes timely completion by one root in a single DAG block and leaves the other five paired counts unchanged.

\subsection{Request-Limit Sensitivity}
\label{app:request-limit}

\noindent\textbf{External arrivals and locked deadlines.}
The Qwen3-8B/910C comparison uses Poisson root arrivals, shared within each workload/rate/block across request limits 64 and 128. Tested rates are 0.24/0.60/1.20 roots/s for ordinary serving, 0.10/0.25/0.50 for loop, and 0.04/0.10/0.20 for MIX7. Each class deadline is $1.5\times$ its independent low-load empirical P95 and is fixed before configuration comparisons. The arrival window is $T=\max(180,24/\lambda)$ seconds; the cohort continues through drain. Confirmation results retain three blocks at each workload, rate, and configuration.

\begin{figure}[!t]
\centering
\includegraphics[width=\columnwidth]{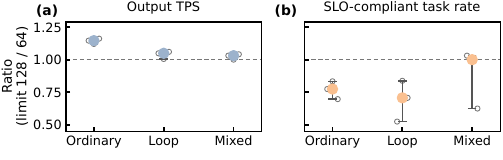}
\vspace{-20pt}
\caption{\textbf{Request-limit sensitivity.} Qwen3-8B/910C, higher/lower limit 128/64. (a) Output TPS ratio. (b) Drain-inclusive timely-task-rate ratio. Highest tested rates are 1.2, 0.5, and 0.2 roots/s for ordinary, Loop, and mixed work. Filled points show paired-block medians, whiskers ranges, and open points individual blocks. Dashed line: parity.}
\label{fig:app-request-limit}
\end{figure}

\begin{figure}[!t]
\centering
\includegraphics[width=\columnwidth]{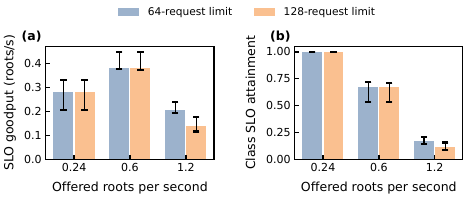}
\vspace{-20pt}
\caption{\textbf{Ordinary-serving load response.} Qwen3-8B/910C, shared arrivals across limits 64/128. (a) Arrival-window SLO goodput. (b) Minimum class attainment. Bars show three confirmation-block medians and ranges under locked deadlines.}
\label{fig:app-slo-ordinary}
\end{figure}

\begin{figure}[!t]
\centering
\includegraphics[width=\columnwidth]{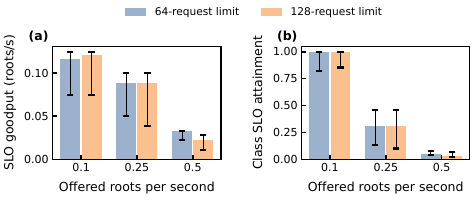}
\vspace{-20pt}
\caption{\textbf{Loop load response.} Same arrival and SLO contract as Figure~\ref{fig:app-slo-ordinary}; each root executes the recorded loop work. Both request limits retain the same per-block arrival stream.}
\label{fig:app-slo-loop}
\end{figure}

\begin{figure}[!t]
\centering
\includegraphics[width=\columnwidth]{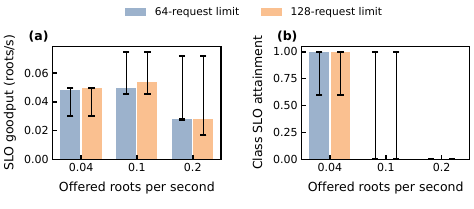}
\vspace{-20pt}
\caption{\textbf{Mixed-workflow load response.} Same contract as Figure~\ref{fig:app-slo-ordinary}, with seven workflow classes and class-specific locked deadlines. Panel (b) reports the least attained class in each confirmation block.}
\label{fig:app-slo-mixed}
\end{figure}

Attainment divides deadline-compliant completed roots by all roots arriving within $T$. Arrival-window cohort goodput divides these timely completions by $T$. Figure~\ref{fig:app-request-limit} instead uses the common per-run observation period $\max(T,T_{\mathrm{cohort}})$ for both output tokens and timely roots. This retains drain in both denominators. Figures~\ref{fig:app-slo-ordinary}--\ref{fig:app-slo-mixed} report arrival-window goodput and the minimum class attainment across the complete load grid.

The admissibility guards require completion of all roots, at most four roots remaining at the end of the arrival window, and nonpositive backlog growth over its second half. A passing rate additionally requires every class in all three confirmation blocks to attain 0.95. The configuration comparisons retain their measured goodput, latency, and drain outcomes.

At the highest tested ordinary-serving load, increasing the request limit from 64 to 128 raises output TPS by 14.6\% but reduces deadline-compliant task rate by 22.5\% (paired-block medians). Loop shows the same direction of change, while the mixed median timely-task-rate ratio is approximately unchanged. This request-limit comparison is distinct from the fixed-limit legal-release/KV study above.

\section{Demand and Fixed-Call Controls}
\label{app:demand}
This appendix supports \S\,\ref{sec:eval-control} with native context and legal release controls first, followed by natural demand, absolute-time trajectories, and aggregate ratios.
\subsection{Legal Optional Control and Native Context Delivery}
\label{app:legal-control}
The four-call binding contains two independent task analyses, synthesis of their artifacts, and verification. Both analysts consume the original task input. Serial control adds one left-to-right control edge while retaining every required data/state edge. Full context delivers each analyst's complete JSON; selected context delivers its \texttt{answer} field through a recorded CPU transformation. Native executions generate the artifacts consumed by later operations. Fixed-work variants retain a native realization's resolved payloads and actual output lengths.

The four-call row of Figure~\ref{fig:eval-release}(b) uses full-context work from three blocks, each containing 12 roots and 48 model calls. DAG enforces original predecessors; Barrier adds the optional analyst edge. The retained four-contract audit also compares Flat (all fixed calls submitted as a non-causal offline reference) and Arrival (observed model-call submission times). Arrival/DAG paired mean-latency ratios range from 0.997 to 1.003. Table~\ref{tab:legal-control-ratios} retains legal-control effects for both context settings.

\begin{table}[!t]
\centering\small
\caption{Serial/parallel ratios at fixed resolved work. Each cell is one paired 12-root block with exact payload and input/output-length checks.}
\label{tab:legal-control-ratios}
\begin{tabular}{@{}llrr@{}}
\toprule
Context & Block & Mean root time & KV integral \\
\midrule
Full & 1 & 1.266 & 0.953 \\
Full & 2 & 1.264 & 0.954 \\
Full & 3 & 1.257 & 0.957 \\
Selected & 1 & 1.280 & 0.958 \\
Selected & 2 & 1.283 & 0.957 \\
Selected & 3 & 1.281 & 0.951 \\
\bottomrule
\end{tabular}
\end{table}

\begin{figure}[!t]
\centering
\includegraphics[width=\columnwidth]{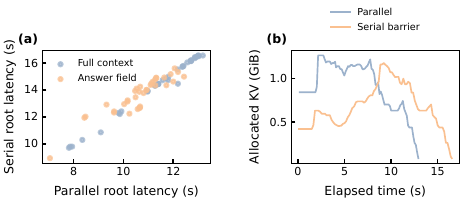}
\vspace{-20pt}
\caption{\textbf{Optional control across context settings.} Qwen3-8B/910C. (a) All 72 paired root observations. (b) Full-context block-1 KV trajectories under parallel/serial control. Each pair retains inputs, output lengths, and required dependencies.}
\label{fig:app-legal-control}
\end{figure}

Native context/control cells use the same 12 parameterized task bindings over three blocks. Table~\ref{tab:native-context-demand} reports generated input work, completed-root latency, and measured KV residency. Native token accounting includes consumed work from every attempt; latency summaries use completed roots. All but one of the 144 native attempts complete, with the retained incomplete-format attempt included in the work ledger. Fixed-work comparisons above use the completed parallel realizations and their preserved request contracts.

\begin{table}[!t]
\centering\scriptsize
\setlength{\tabcolsep}{3pt}
\caption{Native demand of control/context cells: three-block medians of per-root input means and completed-root latency means, and per-cohort KV integrals.}
\label{tab:native-context-demand}
\begin{tabular}{@{}llrrr@{}}
\toprule
Control & Context & Input tokens & Root (s) & KV (GiB\,s) \\
\midrule
Parallel & Full & 1275.4 & 11.318 & 11.522 \\
Serial & Full & 1286.8 & 14.431 & 11.644 \\
Parallel & Selected & 906.7 & 10.753 & 8.585 \\
Serial & Selected & 895.1 & 13.850 & 8.216 \\
\bottomrule
\end{tabular}
\end{table}

\begin{figure}[!t]
\centering
\includegraphics[width=\columnwidth]{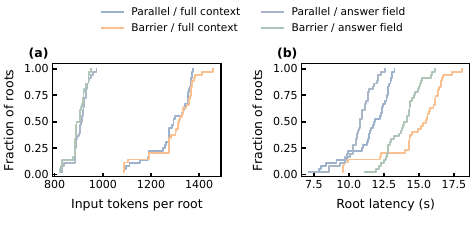}
\vspace{-20pt}
\caption{\textbf{Native control and context delivery.} Qwen3-8B/910C, 12 bindings per cell per block. (a) Generated input tokens per root. (b) Completed-root latency. Native artifacts propagate through synthesis and verification; Table~\ref{tab:native-context-demand} defines the aggregation.}
\label{fig:app-context-delivery}
\end{figure}

\noindent\textbf{Source-workflow native traces.}
We map GPT Researcher's pinned role methods and prompt builders~\cite{gpt_researcher_multiagent} to planning, parallel research/refinement, writing/feedback, and publication stages, with a frozen-document retrieval adapter. Three document bindings cover asyncio, cgroup v2, and HTTP caching. Full delivers each actual reviewed DraftState; Selected retains its draft. Qwen3.5-9B/RTX PRO 5000 pairs use BF16, TP1, request limit 64, prefix caching off, seed 0, and shared 0/0.1/0.2-second arrivals. The six traces reconcile 171 requests, nine exact draft projections, and 32 links from feedback to subsequent prompts. Three paired serving blocks reuse the three Full captures (83 calls), their 0/0.1/0.2-second arrivals, snapshots, and recorded output lengths. DAG retains required dependencies; Barrier adds optional ordering between the three research branches. Each pair uses the same physical GPU and empty initial KV/queues, with DAG/Barrier order AB/BA/AB across blocks.

\begin{table}[!t]
\centering\scriptsize
\setlength{\tabcolsep}{2.5pt}
\caption{Source-workflow native work underlying Figure~\ref{fig:eval-release}(a). Entries are Full/Selected, one native pair per document binding. Token counts are in thousands.}
\label{tab:source-native-work}
\begin{tabular}{@{}lrrr@{}}
\toprule
Binding & Calls & Input (k) & Output (k) \\
\midrule
Asyncio & 28/28 & 77.038/73.496 & 19.589/18.122 \\
cgroup v2 & 30/30 & 85.928/77.785 & 24.324/24.785 \\
HTTP cache & 25/30 & 68.495/87.922 & 16.154/23.131 \\
\bottomrule
\end{tabular}
\end{table}

\noindent\textbf{Generated graph structure.}
Table~\ref{tab:source-native-structural-diff} decomposes the six native traces by their realized review paths. All activate three research branches and eight stages. In the HTTP pair, the additional section revision contributes two model calls (revision and review), while an additional writing pass contributes three (writing, headers, and checking). The LLM graph consequently grows from 25 to 30 calls and from 15 to 20 calls on its longest path. Native execution regenerates research artifacts and feedback decisions in each arm; the table separates their research and writing contributions.

\begin{table}[!t]
\centering\scriptsize
\setlength{\tabcolsep}{2.5pt}
\caption{Native structural trace differences. Entries are Full/Selected, one native pair per binding. Refinements count section-draft revisions; writing passes include the initial pass. Edges and depth use the LLM-only graph after collapsing non-LLM paths and removing transitive edges. Depth counts calls.}
\label{tab:source-native-structural-diff}
\begin{tabular}{@{}lrrrrr@{}}
\toprule
Binding & Calls & \shortstack{Refine-\\ments} & \shortstack{Writing\\passes} & Edges & Depth \\
\midrule
Asyncio & 28/28 & 3/3 & 3/3 & 29/29 & 18/18 \\
cgroup v2 & 30/30 & 4/4 & 3/3 & 31/31 & 20/20 \\
HTTP cache & 25/30 & 3/4 & 2/3 & 26/31 & 15/20 \\
\bottomrule
\end{tabular}
\end{table}

\noindent\textbf{Frozen-graph representation control.}
For each Full trace, we project the first writer's actual recorded DraftStates to their draft fields and retain its operation set and required edges. This offline prompt edit preserves 28, 30, and 25 model calls for Asyncio, cgroup v2, and HTTP, respectively; it records structural properties only. The HTTP Selected graph's additional writing/checking cycle is produced by executing the collaboration logic with generated outputs. XPerf supports trace collection, DAG construction, workload synthesis, and dependency-aware and token-wise replay~\cite{xperf}. The control here describes a fixed observed graph, while MAServeBench's native context configuration generates the work that can subsequently be frozen for legal release and resource comparisons.

\subsection{Natural Demand and Fixed-Call Trajectories}
\begin{figure}[!t]
\centering
\includegraphics[width=\columnwidth]{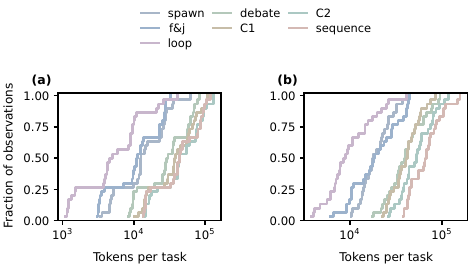}
\vspace{-20pt}
\caption{Natural input/output demand across 210 task/structure executions. Shared bindings retain generated artifacts and output lengths; model-call medians range from two for loop to sixteen for the composed sequence.}
\label{fig:app-natural}
\end{figure}

\begin{figure}[!t]
\centering
\includegraphics[width=\columnwidth]{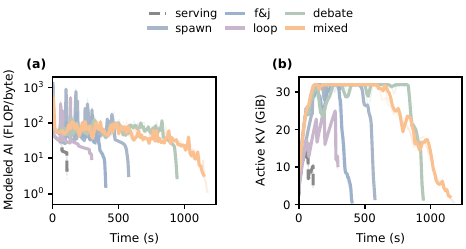}
\vspace{-20pt}
\caption{Absolute-time demand on Qwen3-8B/910C for 64 users. Every block is plotted; block zero is emphasized consistently. Curves end at their own run completion.}
\label{fig:app-absolute}
\end{figure}

\begin{figure}[!t]
\centering
\includegraphics[width=\columnwidth]{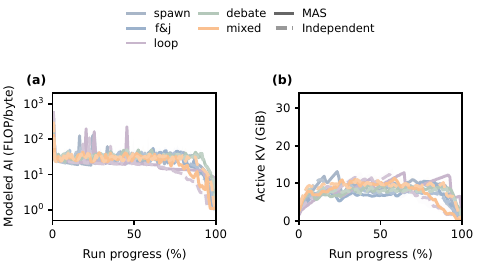}
\vspace{-20pt}
\caption{Fixed-call demand controls on Qwen3-8B/910C, request limit 32. Solid lines preserve workflow dependencies; dashed lines release independent calls from the same resolved work set. Each condition contains three blocks. Horizontal axes show each cohort's own progress.}
\label{fig:app-calls32}
\end{figure}

\begin{figure}[!t]
\centering
\includegraphics[width=\columnwidth]{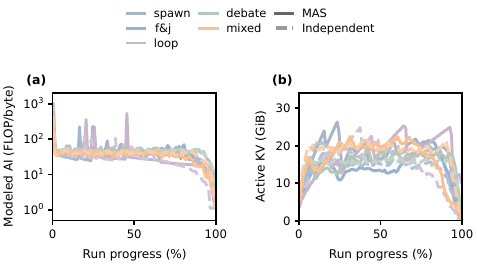}
\vspace{-20pt}
\caption{Fixed-call demand controls on Qwen3-8B/910C, request limit 64. Solid lines preserve workflow dependencies; dashed lines release independent calls from the same resolved work set. Each condition contains three blocks. Horizontal axes show each cohort's own progress.}
\label{fig:app-calls64}
\end{figure}

\begin{figure}[!t]
\centering
\includegraphics[width=\columnwidth]{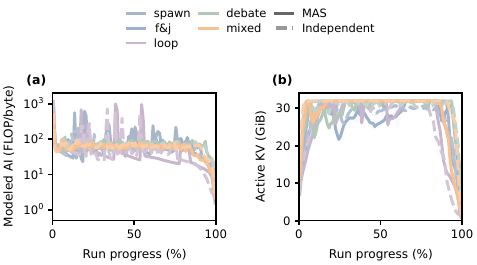}
\vspace{-20pt}
\caption{Fixed-call demand controls on Qwen3-8B/910C, request limit 128. Solid lines preserve workflow dependencies; dashed lines release independent calls from the same resolved work set. Each condition contains three blocks. Horizontal axes show each cohort's own progress.}
\label{fig:app-calls128}
\end{figure}

The natural demand set contains 30 task bindings for each of seven structures. Figure~\ref{fig:app-natural} gives their input/output distributions. The primary full-attention comparison contains 54 windows: three user counts, six serving/workflow conditions, and three blocks. Figure~\ref{fig:app-absolute} shows their absolute-time behavior for 64 users. The complementary fixed-call study contains 45 paired comparisons: five workflow conditions, three request limits, and three blocks. Both arms preserve the resolved call set; dependency release determines its temporal organization. Figures~\ref{fig:app-calls32}--\ref{fig:app-calls128} retain the complete trajectories.

\begin{figure}[!t]
\centering
\includegraphics[width=\columnwidth]{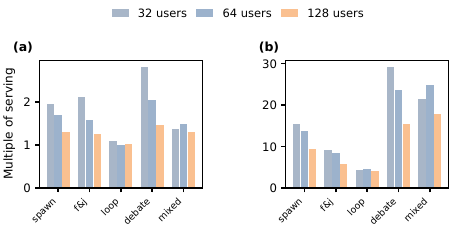}
\vspace{-20pt}
\caption{\textbf{Demand relative to serving.} Matched user concurrency. Three-block median ratios: (a) aggregate $\sum F/\sum D$; (b) KV residency integral $\int K(t)\,dt$.}
\label{fig:app-demand-ratios}
\end{figure}

\noindent\textbf{Arithmetic-intensity accounting.}
For Qwen3-8B, let $L=36$, hidden width $H=4096$, intermediate width $I=12288$, query-head count $N_h=32$, KV-head count $N_k=8$, head width $d=128$, vocabulary size $V$ from the model configuration, and element size $s=2$ bytes. Projection parameters are
\[
P=L[H(N_hd+2N_kd)+N_hdH+3HI].
\]
At a scheduler step, request $i$ has $c_i$ computed tokens and $q_i$ scheduled tokens. Let $T=\sum_iq_i$, $A=\sum_i[q_ic_i+q_i(q_i+1)/2]$, and let $r$ count requests requiring output-head evaluation. Modeled work is
\[
F=2PT+4LN_hdA+2HVr.
\]
The byte model sums streaming weights $s(P+HV\mathbf{1}_{r>0})$, logical KV reads $2LN_kds\sum_i(c_i+q_i)$, KV writes $2LN_kdsT$, and activation traffic $2sLT(4H+3I+2N_kd)$. Intensity is $F/D$. This accounting supplies a common logical-demand definition. Physical cache measurements appear in \S\,\ref{sec:eval-capacity}. Aggregate run intensity uses $\sum F/\sum D$.

Allocated KV occupancy (GiB) is scheduler block utilization multiplied by the allocated KV pool, including block rounding. The time-integrated allocated KV footprint, $\int K(t)\,dt$ in GiB\,$\cdot$\,s, accumulates this occupancy over cohort wall time. Curves use duration-weighted values in 100 progress bins, with three-block medians and ranges. These ranges describe block variation. Natural demand, fixed-call controls, and same-user demand retain their respective work definitions.

\section{Compute Geometry and Implementation Checks}
\label{app:compute}
\begin{figure}[!t]
\centering
\includegraphics[width=\columnwidth]{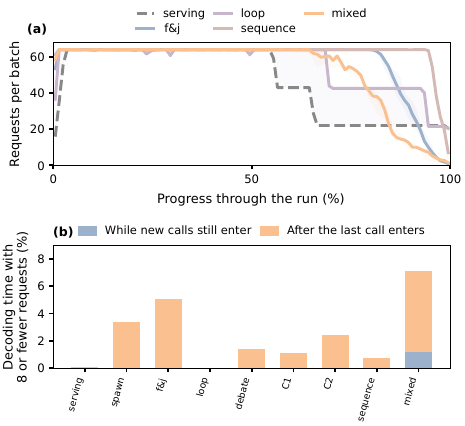}
\vspace{-20pt}
\caption{Batch evolution and final drain, Qwen3.5-9B/RTX PRO 5000, 64 users and a 64-request limit. (b) decoding time with at most eight requests, split at the first scheduling of the last new model call. Final drain accounts for most low-batch exposure.}
\label{fig:app-drain}
\end{figure}

This appendix supports \S\,\ref{sec:eval-hardware} with complete projection timings, application shape coverage, and emitted gather-path checks.
Table~\ref{tab:projection-support} records the seven projection families used to connect application demand to GPU controls. One-weight and rotating-eight-weight cases probe local reuse and multiple addresses at each controlled row count. The additional 512-row cases in Table~\ref{tab:wide-controls} extend the shape comparison. Figure~\ref{fig:app-drain} separates the full batch trajectory from low-batch exposure before and after final admission.

\begin{table}[!t]
\centering\scriptsize
\setlength{\tabcolsep}{3pt}
\caption{Qwen3.5-9B projection support. Footprints count distinct BF16 weight instances in the model.}
\label{tab:projection-support}
\begin{tabular}{@{}lrrrr@{}}
\toprule
Family & $N$ & $K$ & Instances & MiB/weight \\
\midrule
Auxiliary & 64 & 4096 & 24 & 0.5 \\
Output & 4096 & 4096 & 32 & 32 \\
Down & 4096 & 12288 & 32 & 96 \\
Attention input & 10240 & 4096 & 8 & 80 \\
Recurrent input & 12288 & 4096 & 24 & 96 \\
Gate/up & 24576 & 4096 & 32 & 192 \\
Head & 248320 & 4096 & 1 & 1940 \\
\bottomrule
\end{tabular}
\end{table}

\begin{table}[!t]
\centering\scriptsize
\setlength{\tabcolsep}{3pt}
\caption{Wider-row GPU controls: CUDA-event time in $\mu$s, with one reused weight.}
\label{tab:wide-controls}
\begin{tabular}{@{}rrrr@{}}
\toprule
$M$ & $N$ & A800 & RTX PRO 5000 \\
\midrule
512 & 4096 & 90.83 & 121.85 \\
512 & 24576 & 447.50 & 549.53 \\
\bottomrule
\end{tabular}
\end{table}

\begin{table}[!t]
\centering\scriptsize
\setlength{\tabcolsep}{3pt}
\caption{Matched normalization: native Memory-pass durations in $\mu$s.}
\label{tab:norm-controls}
\begin{tabular}{@{}lrrr@{}}
\toprule
Rows & 910C & 950DT & Ratio \\
\midrule
1 & 4.397 & 2.007 & 2.19 \\
8 & 3.445 & 2.867 & 1.20 \\
32 & 5.480 & 5.017 & 1.09 \\
\bottomrule
\end{tabular}
\end{table}

\noindent\textbf{Native workflow-to-matrix attribution.}
The 18 Qwen3-8B/910C captures in Figure~\ref{fig:eval-execution}(b) contain 380,480 matrix calls, totaling 48.890863 seconds of kernel time. Each of the nine workflow/serving conditions has one Early and one Late capture at 64 roots and a 128-request limit. Two-dimensional input/output tensor records provide $M,N,K$ for all captured matrix calls. $M$ denotes operator-visible tensor rows, including implementation padding. Call shares use invocation counts; time shares sum kernel durations within each capture. Table~\ref{tab:native-matrix-rows} retains all nine conditions.

Explicit profiler flows connect matrix kernels through batch markers and scheduler request IDs to workflow/stage membership. These joins cover 17.5\% of pooled matrix time; 10.3\% belongs to batches with one workflow/template-stage identity, and shared batches retain their joint membership in the ledger. Scheduler token state identifies pure-decode or prefill/mixed batches. In C2's Late capture, all 1,008 matrix calls with $M>128$ map to the Debate stage: 1.186310 seconds, or 44.8\% of that capture's matrix time. This traces a concrete collaboration stage through scheduling to its executed geometry.

\begin{table}[!t]
\centering\scriptsize
\setlength{\tabcolsep}{3pt}
\caption{Native matrix rows, Qwen3-8B/910C. Shares are percentages of each capture's matrix kernel time. Early/Late are capture windows; the last column gives Late time at $M>128$.}
\label{tab:native-matrix-rows}
\begin{tabular}{@{}lrrrrr@{}}
\toprule
& \multicolumn{2}{c}{Early} & \multicolumn{2}{c}{Late} & Late \\
Workflow & Modal $M$ & Share & Modal $M$ & Share & $M>128$ \\
\midrule
Serving & 64 & 100.0 & 24 & 92.5 & 0.0 \\
Spawn & 64 & 100.0 & 128 & 84.0 & 15.6 \\
Fork--Join & 128 & 100.0 & 128 & 66.9 & 32.7 \\
Loop & 64 & 100.0 & 64 & 100.0 & 0.0 \\
Debate & 128 & 100.0 & 128 & 89.9 & 9.8 \\
C1 & 64 & 100.0 & 96 & 52.2 & 41.6 \\
C2 & 128 & 100.0 & 88 & 51.2 & 44.8 \\
Sequence & 128 & 100.0 & 120 & 57.8 & 36.8 \\
Mixed & 128 & 100.0 & 96 & 51.6 & 41.5 \\
\bottomrule
\end{tabular}
\end{table}

\noindent\textbf{Application geometry coverage.}
Across 12 sampled 910C hybrid-model captures, the three NPU matrix reference families in Figure~\ref{fig:eval-npu-compute}(a), with $K=4096$ and $N\in\{64,4096,24576\}$, cover 23.222 of 42.322 seconds of sampled matrix time by $N,K$ (54.9\%). Including the reference row counts $M\in\{1,8,32\}$ gives complete $M,N,K$ coverage of approximately 51.4\%. Both shares use the pooled sampled matrix kernel time as their denominator. Full-attention batch-64 captures match approximately 53.7\% by $N,K$; their native row count is 64. The 950DT hybrid captures match approximately 45.3\% by $N,K$ and include native 40-row shapes. These coverage figures connect controlled demands to their application source. Native application shape pairs remain available in the archived evidence tables.

The 950DT compile comparison uses four exact source/index pairs. Binary and kernel inspection distinguish SIMD from SIMT; the hybrid-template binary matches SIMD. The separate library gather references provide an additional implementation choice. This implementation-level evidence motivates path selection for auxiliary work.

\begin{table}[!t]
\centering\footnotesize
\setlength{\tabcolsep}{2pt}
\caption{All 28 GPU projection event-time cases for Figure~\ref{fig:eval-gpu-compute}. Times are microseconds; ratio is A800/RTX PRO 5000. Weight count is the rotating address set size.}
\label{tab:gpu-all-timings}
\begin{tabular}{lrrrrr}
\toprule
Family & Rows & Weights & A800 & RTX PRO & Ratio \\
\midrule
Auxiliary & 1 & 1 & 6.696 & 2.964 & 2.259 \\
Auxiliary & 1 & 8 & 6.528 & 2.960 & 2.205 \\
Output & 1 & 1 & 23.312 & 13.089 & 1.781 \\
Output & 1 & 8 & 25.136 & 37.601 & 0.668 \\
Down & 1 & 1 & 67.032 & 37.980 & 1.765 \\
Down & 1 & 8 & 66.832 & 109.479 & 0.610 \\
Attention & 1 & 1 & 60.808 & 28.061 & 2.167 \\
Attention & 1 & 8 & 55.920 & 90.146 & 0.620 \\
Recurrent & 1 & 1 & 69.432 & 39.379 & 1.763 \\
Recurrent & 1 & 8 & 69.040 & 107.728 & 0.641 \\
Gate/up & 1 & 1 & 129.512 & 212.328 & 0.610 \\
Gate/up & 1 & 8 & 129.384 & 212.352 & 0.609 \\
Head & 1 & 1 & 1128.072 & 2055.631 & 0.549 \\
Head & 1 & 8 & 1128.192 & 2055.775 & 0.549 \\
Auxiliary & 8 & 1 & 7.184 & 3.739 & 1.921 \\
Auxiliary & 8 & 8 & 7.160 & 3.754 & 1.907 \\
Output & 8 & 1 & 23.600 & 21.607 & 1.092 \\
Output & 8 & 8 & 25.512 & 40.371 & 0.632 \\
Down & 8 & 1 & 67.208 & 76.406 & 0.880 \\
Down & 8 & 8 & 67.392 & 118.172 & 0.570 \\
Attention & 8 & 1 & 60.152 & 33.033 & 1.821 \\
Attention & 8 & 8 & 56.840 & 91.289 & 0.623 \\
Recurrent & 8 & 1 & 65.984 & 55.174 & 1.196 \\
Recurrent & 8 & 8 & 65.992 & 108.613 & 0.608 \\
Gate/up & 8 & 1 & 137.448 & 212.245 & 0.648 \\
Gate/up & 8 & 8 & 126.048 & 211.617 & 0.596 \\
Head & 8 & 1 & 1288.656 & 2073.283 & 0.622 \\
Head & 8 & 8 & 1169.336 & 2073.343 & 0.564 \\
\bottomrule
\end{tabular}
\end{table}

\begin{table}[!t]
\centering\footnotesize
\setlength{\tabcolsep}{2pt}
\caption{All 12 gather path checks for Figure~\ref{fig:eval-npu-compute}. Within each layout, source/index hashes and the numerical contract match; binary hashes identify the actual compiled path. Times are native Memory-pass microseconds.}
\label{tab:gather-binary}
\begin{tabular}{rlllrl}
\toprule
Bytes & Layout & Template & Path & Time & Binary SHA \\
\midrule
64 & ordered & SIMD & SIMD & 24.608 & \texttt{377f67f79470} \\
64 & ordered & Hybrid & SIMD & 24.660 & \texttt{377f67f79470} \\
64 & ordered & SIMT & SIMT & 3.566 & \texttt{01a1d6e8b3e6} \\
64 & permuted & SIMD & SIMD & 24.475 & \texttt{377f67f79470} \\
64 & permuted & Hybrid & SIMD & 24.507 & \texttt{377f67f79470} \\
64 & permuted & SIMT & SIMT & 3.724 & \texttt{01a1d6e8b3e6} \\
512 & ordered & SIMD & SIMD & 23.299 & \texttt{377f67f79470} \\
512 & ordered & Hybrid & SIMD & 23.073 & \texttt{377f67f79470} \\
512 & ordered & SIMT & SIMT & 3.578 & \texttt{01a1d6e8b3e6} \\
512 & permuted & SIMD & SIMD & 23.205 & \texttt{377f67f79470} \\
512 & permuted & Hybrid & SIMD & 22.994 & \texttt{377f67f79470} \\
512 & permuted & SIMT & SIMT & 3.585 & \texttt{01a1d6e8b3e6} \\
\bottomrule
\end{tabular}
\end{table}

\section{Physical Data Supply, Attention, and State Events}
\label{app:memory}
\begin{figure}[!t]
\centering
\includegraphics[width=\columnwidth]{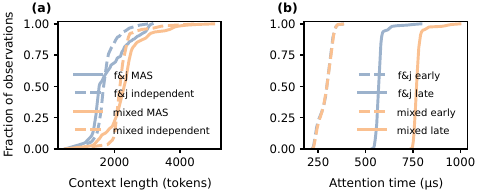}
\vspace{-20pt}
\caption{Full-attention context and native execution distributions, Qwen3-8B/910C, 64 users with a 64-request limit. The four context controls include every pure-decoding step, weighted by step duration. Panel (b) contains four MAS early/late captures; all eight arms/phases contain 53,172 native attention calls.}
\label{fig:app-context}
\end{figure}

This appendix supports \S\,\ref{sec:eval-capacity}: physical projection reads, exact-context attention controls, and allocated-pool state events retain their own operator and run records.
Eight representative batch/context samples are linked to the original scheduler by batch key, row count, and exact context multiset. Each full/half pair fixes backing allocation and changes the context vector; Table~\ref{tab:context-controls} reports all pairs. Application attention is FusedInferAttentionScore with Qwen3-8B geometry. The separate cross-device controls use batch 32 with fixed 512\,MiB backing. Figure~\ref{fig:app-context} retains the time-weighted batch-mean context distributions and native attention distributions. The interval 1024--4096 covers 95.8--98.7\% of pure-decoding step-mean-context time in these controls.

\begin{table}[!t]
\centering\scriptsize
\setlength{\tabcolsep}{3pt}
\caption{Application-derived context controls, batch 64. Operator times are in $\mu$s. Quantiles select representative context samples within each arm.}
\label{tab:context-controls}
\begin{tabular}{@{}llrrrr@{}}
\toprule
Workflow & Arm & Quantile & Full & Half & Ratio \\
\midrule
Mixed & MAS & 0.25 & 551.8 & 288.0 & 1.92 \\
Mixed & MAS & 0.75 & 743.4 & 383.3 & 1.94 \\
Mixed & independent & 0.25 & 691.1 & 362.1 & 1.91 \\
Mixed & independent & 0.75 & 742.4 & 376.3 & 1.97 \\
f\&j & MAS & 0.25 & 415.7 & 207.8 & 2.00 \\
f\&j & MAS & 0.75 & 512.0 & 263.0 & 1.95 \\
f\&j & independent & 0.25 & 571.3 & 283.5 & 2.01 \\
f\&j & independent & 0.75 & 601.7 & 304.5 & 1.98 \\
\bottomrule
\end{tabular}
\end{table}

The state-pressure comparison uses 64 users and a 128-request limit. Mixed blocks record 431--471 preemptions each and a prefill processing factor of 1.76--1.84 relative to original input. Loop-only blocks record zero preemptions and a factor of one. Victim IDs link the event stream to request residence and workflow membership. Joining all 1,372 native preemption events to response IDs and the common execution clock places them between first scheduling and completion. At these events, 90--128 submitted unfinished requests span five to seven workflow classes; each event falls in a 1\%-duration bin whose time-weighted mean KV allocation is at least 95\%. Submitted unfinished intervals include request waiting and pauses. This joins collaboration-shaped exposure, allocated state, and preemption at their recorded time resolution. Allocated-pool occupancy, logical access, and physical DRAM reads retain their separate units.

\begin{table}[!t]
\centering\footnotesize
\setlength{\tabcolsep}{1.6pt}
\caption{All physical reads underlying Figure~\ref{fig:eval-gpu-memory}. Values are MiB per complete sampled primitive call; one-weight size is the plotting denominator. Event timing is retained in a separate pass in Table~\ref{tab:gpu-all-timings}.}
\label{tab:gpu-physical-reads}
\begin{tabular}{lrrrrr}
\toprule
Family & Rows & Weights & \shortstack{One\\weight} & \shortstack{A800\\reads} & \shortstack{RTX PRO\\reads} \\
\midrule
Auxiliary & 1 & 1 & 0.500 & 0.000000 & 0.000000 \\
Auxiliary & 1 & 8 & 0.500 & 0.000000 & 0.000000 \\
Output & 1 & 1 & 32.000 & 25.621948 & 0.000000 \\
Output & 1 & 8 & 32.000 & 32.013550 & 31.997192 \\
Down & 1 & 1 & 96.000 & 96.031555 & 60.832153 \\
Down & 1 & 8 & 96.000 & 96.031921 & 96.000488 \\
Attention & 1 & 1 & 80.000 & 80.015686 & 5.151978 \\
Attention & 1 & 8 & 80.000 & 80.015747 & 80.001465 \\
Recurrent & 1 & 1 & 96.000 & 96.019836 & 68.392090 \\
Recurrent & 1 & 8 & 96.000 & 96.020691 & 96.003784 \\
Gate/up & 1 & 1 & 192.000 & 192.063843 & 192.018433 \\
Gate/up & 1 & 8 & 192.000 & 192.063721 & 192.002197 \\
Head & 1 & 1 & 1940.000 & 1940.062195 & 1940.319824 \\
Head & 1 & 8 & 1940.000 & 1940.057068 & 1940.034668 \\
Auxiliary & 8 & 1 & 0.500 & 0.000000 & 0.000000 \\
Auxiliary & 8 & 8 & 0.500 & 0.000000 & 0.000000 \\
Output & 8 & 1 & 32.000 & 25.926270 & 0.000000 \\
Output & 8 & 8 & 32.000 & 32.058838 & 32.021606 \\
Down & 8 & 1 & 96.000 & 96.195557 & 48.296997 \\
Down & 8 & 8 & 96.000 & 96.195862 & 96.130859 \\
Attention & 8 & 1 & 80.000 & 80.070251 & 0.359863 \\
Attention & 8 & 8 & 80.000 & 80.070801 & 80.009766 \\
Recurrent & 8 & 1 & 96.000 & 96.070496 & 39.539429 \\
Recurrent & 8 & 8 & 96.000 & 96.070740 & 96.009521 \\
Gate/up & 8 & 1 & 192.000 & 192.075806 & 192.011963 \\
Gate/up & 8 & 8 & 192.000 & 192.072998 & 192.011719 \\
Head & 8 & 1 & 1940.000 & 1940.596619 & 1940.072144 \\
Head & 8 & 8 & 1940.000 & 1940.673828 & 1940.149658 \\
\bottomrule
\end{tabular}
\end{table}

\begin{table*}[!t]
\centering\scriptsize
\setlength{\tabcolsep}{3pt}
\caption{State-event and repeated-work ledger for Figure~\ref{fig:eval-state}. Mixed and loop-only controls use 64 users, request limit 128, and the same 31.99\,GiB allocated pool. Victims count distinct mixed-run request IDs.}
\label{tab:state-event-ledger}
\begin{tabular}{rrrrrrrr}
\toprule
Block & Preemptions & Victims & Processed prefill & Source input & Factor & Loop evictions & Loop factor \\
\midrule
0 & 470 & 186 & 1052422 & 571973 & 1.840 & 0 & 1.000 \\
1 & 471 & 189 & 1025513 & 570441 & 1.798 & 0 & 1.000 \\
2 & 431 & 180 & 1004811 & 571822 & 1.757 & 0 & 1.000 \\
\bottomrule
\end{tabular}
\end{table*}

\subsection{Fixed-Work KV-Budget Intervention}
\label{app:kv-budget}
\begin{figure}[!t]
\centering
\includegraphics[width=\columnwidth]{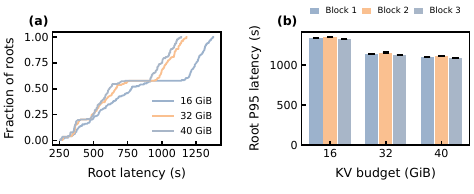}
\vspace{-20pt}
\caption{\textbf{Root outcomes across KV budgets.} Same fixed-work comparison as Figure~\ref{fig:eval-budget}. (a) All 192 root latencies per budget across three blocks. (b) Each block's root P95.}
\label{fig:app-budget-roots}
\end{figure}

\begin{figure}[!t]
\centering
\includegraphics[width=\columnwidth]{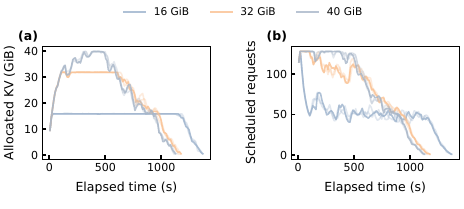}
\vspace{-20pt}
\caption{\textbf{State and batch trajectories across KV budgets.} Same Qwen3-8B/910C comparison as Figure~\ref{fig:eval-budget}. All nine runs are shown over absolute time; every trajectory ends at its own completion.}
\label{fig:app-budget-trajectories}
\end{figure}

The budget comparison in Figure~\ref{fig:eval-budget} fixes a seven-workflow cohort of 64 roots and a 128-request limit on Qwen3-8B/910C. Each block uses identical root arrivals, resolved payloads, input/output lengths, and required dependencies across the three budgets. Execution order alternates across blocks. All 576 attempted roots complete. Table~\ref{tab:kv-budget} retains each run's actual allocator storage, preemptions, processed/source prefill, and root P95. Figures~\ref{fig:app-budget-roots} and~\ref{fig:app-budget-trajectories} retain root distributions and absolute-time state/batch trajectories.

For this budget study, the original estimator takes the sorted latency at zero-based index $\lceil .95(n-1)\rceil$ (index 60 for 64 roots). Within-block 40/16\,GiB prefill-factor ratios are 0.470470, 0.480313, and 0.490570; root-P95 ratios are 0.822051, 0.825358, and 0.825079. Their medians imply reductions of 51.97\% and 17.49\%. Both quantities decrease monotonically from 16 through 32 to 40\,GiB in every block. These fixed-work windows are independent of the native mixed-cohort observations above.

\begin{table}[!t]
\centering\scriptsize
\setlength{\tabcolsep}{3pt}
\caption{KV-budget run ledger. Actual allocation is GiB; factor is processed/source prefill; P95 is seconds. Every row contains 64 completed roots.}
\label{tab:kv-budget}
\begin{tabular}{@{}rrrrrr@{}}
\toprule
Block & Budget & Actual & Preemptions & Factor & P95 \\
\midrule
1 & 16 & 15.996 & 887 & 2.523 & 1344.587 \\
2 & 16 & 15.996 & 899 & 2.557 & 1347.172 \\
3 & 16 & 15.996 & 900 & 2.478 & 1323.505 \\
1 & 32 & 31.992 & 460 & 1.843 & 1141.455 \\
2 & 32 & 31.992 & 511 & 1.933 & 1157.747 \\
3 & 32 & 31.992 & 425 & 1.791 & 1130.362 \\
1 & 40 & 39.990 & 96 & 1.187 & 1105.319 \\
2 & 40 & 39.990 & 134 & 1.228 & 1111.899 \\
3 & 40 & 39.990 & 120 & 1.216 & 1091.996 \\
\bottomrule
\end{tabular}
\end{table}

\section{Stage and Complete-task Attribution}
\label{app:interference}
This appendix supports \S\,\ref{sec:eval-interference} with the fixed-work temporal-mixing audit first, followed by background composition, cohort characterization, and earlier sensitivity results.
\subsection{Balanced Fixed-Population Temporal Mixing}
\label{app:balanced-temporal-mixing}
Figure~\ref{fig:eval-temporal-mixing} uses nine tagged Loop targets (three each for GSM8K, PubMedQA, and Science) and a background of 19 Loop, 18 Spawn, and 18 Fork--Join roots in each of three paired blocks. Three recorded task bindings are assigned cyclically within each background workflow, with a block rotation fixed before execution. Both arms replay the same complete trace multiset, resolved inputs, output lengths, required dependencies, tool delays, and targets. Target arrivals are 0--0.16\,s in 0.02-second steps. The 55 background slots are 2, 4, \ldots, 110\,s. Lexicographic workflow order and stable root IDs within each class determine round-robin interleaving (A) or class concatenation (B). Only the mapping of background traces to those slots changes.

All pairs use Qwen3-8B/910C, BF16, TP1, request limit 128, the same 31.99\,GiB KV pool, prefix caching off, and length-locked replay. Execution order is A/B, B/A, A/B across blocks; each pair uses the same physical card and backend process, with zero KV allocation and empty queues verified before each arm. Table~\ref{tab:balanced-mixing-audit} reports realized totals, equal within every pair. All six arms complete 64 roots and 236 calls; no failures or KV preemptions occur. Canonical request hashes, per-operation dependencies, global slots, target identities and arrivals, and realized token counts reconcile with the locked manifest and execution ledgers.

Panel (a) weights actual scheduler intervals by elapsed time within the common 2--110\,s arrival window. Its denominator sums intervals containing at least one target request. The Loop, Spawn, and Fork--Join rows retain intervals also containing at least one request from that background workflow; All three requires all three background workflows simultaneously. These overlapping presence shares use the same denominator. Multiple targets in one interval contribute once. An interval ends at the next scheduling start, capped by the latest completion among its scheduled requests. Every scheduler request ID maps to the request ledger, and first-scheduling timestamps reconcile with that mapping. The reported shares summarize elapsed scheduled intervals rather than request counts. Pure Decode batch size, weighted by interval duration over the target lifetime, has three-block medians of 79.8 (A) and 72.9 (B).

For panel (b), scheduled residence is request completion minus first scheduling, including subsequent pauses. The immutable dependency graph identifies the first and second LLM calls; the second depends on the first. Each point estimate first averages the three paired A-minus-B differences within one binding and block, then takes the median and actual range across the three blocks. For panel (c), each block's percentage is $100(\overline{L}_A/\overline{L}_B-1)$, with arithmetic root-latency means over three targets per binding or all nine targets for Overall. The displayed median and range summarize those three ratios. Overall is recomputed from the nine roots, and the two call medians are summarized separately; neither statistic is formed by adding binding or call summaries. The released evidence retains all 27 root pairs and 54 call pairs, including operation IDs, arrival, ready, first-scheduling, and completion timestamps. Both request completion times and root outcomes reconcile with the original timestamp logs.

Table~\ref{tab:balanced-mixing-paths} retains every binding/block result and overall path component. All 27 targets have shorter first-call and longer second-call residence. The latter exceeds the former in GSM8K and PubMedQA but only partly offsets it in Science. Small overall pre-submission and backend-waiting differences leave the opposing residence changes as the main contributions to the binding-specific root outcomes.

\begin{table}[!t]
\centering\scriptsize
\setlength{\tabcolsep}{2.2pt}
\caption{Balanced temporal-mixing audit. A and B have identical per-block totals. The common background is 19 Loop + 18 Spawn + 18 Fork--Join roots; each arm also retains nine targets. Shares are time-weighted within 2--110\,s.}
\label{tab:balanced-mixing-audit}
\begin{tabular}{@{}rrrrrr@{}}
\toprule
Block & Calls & Input & Output & A share & B share \\
\midrule
1 & 236 & 180,249 & 517,259 & 96.278\% & 31.474\% \\
2 & 236 & 179,890 & 516,780 & 96.275\% & 31.442\% \\
3 & 236 & 180,229 & 515,145 & 96.280\% & 31.461\% \\
\bottomrule
\end{tabular}
\end{table}

\begin{table}[!t]
\centering\scriptsize
\setlength{\tabcolsep}{2.0pt}
\caption{Balanced temporal mixing, all three paired blocks. Top: three targets per binding/block; call entries are mean paired scheduled-residence differences (A minus B, seconds). Root entries are ratios of means, not means of individual ratios. Bottom: overall mean path differences over nine targets. Pre/Wait are milliseconds; residence and root differences are seconds.}
\label{tab:balanced-mixing-paths}
\begin{tabular}{@{}llrrr@{}}
\toprule
Binding & Block & Call 1 (s) & Call 2 (s) & Root change \\
\midrule
GSM8K & 1 & -24.027 & +31.264 & +2.169\% \\
GSM8K & 2 & -21.461 & +32.747 & +3.439\% \\
GSM8K & 3 & -23.953 & +28.384 & +1.351\% \\
PubMedQA & 1 & -17.206 & +23.613 & +2.136\% \\
PubMedQA & 2 & -14.452 & +24.905 & +3.525\% \\
PubMedQA & 3 & -16.170 & +19.943 & +1.253\% \\
Science & 1 & -13.162 & +5.712 & -3.752\% \\
Science & 2 & -10.506 & +4.862 & -2.892\% \\
Science & 3 & -11.960 & +2.899 & -4.554\% \\
\midrule
Block & $\Delta$Pre (ms) & $\Delta$Wait (ms) & $\Delta$Res (s) & $\Delta$Root (s) \\
\midrule
1 & -0.029 & +2.911 & +2.065 & +2.069 \\
2 & +0.008 & -18.914 & +5.365 & +5.346 \\
3 & -0.014 & +18.477 & -0.286 & -0.268 \\
\bottomrule
\end{tabular}
\end{table}

\subsection{Background Composition and Demand}
\label{app:matched-background}
The Qwen3-8B/910C study in Figure~\ref{fig:eval-cotenancy} uses BF16, tensor parallelism one, and a single serving instance. Each arm retains nine tagged Loop targets and 55 background roots, with identical root IDs and arrivals within each paired block. The baseline has 55 Loop backgrounds; the perturbed arm retains 53 and substitutes one Spawn and one Fork--Join for the other two. Both arms preserve the targets' canonical requests, input and output lengths, dependencies, and bindings. The request limit is 128, prefix caching is disabled, and the model-length and batched-token limits are both 8192.

Table~\ref{tab:matched-target-background} gives background work and target outcomes for all six windows. Background input/output length distributions have Kolmogorov--Smirnov distances of approximately 0.026/0.043. Inputs rise by 4.45--4.48\%, outputs by 4.78--4.82\%, and calls by 5.45\%; the experiment therefore describes a limited composition perturbation with approximately matched work. Each block contains three repetitions of each frozen input: \texttt{GSM8K-0-1}, \texttt{PUBMEDQA-0-1}, and \texttt{SCIENCE-0-1}. The 27 paired target executions yield 54 paired model requests, comprising repeated executions of these three inputs.

\begin{table*}[!t]
\centering\scriptsize
\setlength{\tabcolsep}{3pt}
\caption{Near-matched background work and fixed-target outcomes. Each window contains nine targets and 55 background roots. Calls and tokens refer only to the background; mean/P95 are target-root latency in seconds.}
\label{tab:matched-target-background}
\begin{tabular}{@{}llrrrrr@{}}
\toprule
Block & Background & Calls & Input tokens & Output tokens & Target mean & Target P95 \\
\midrule
1 & Loop only & 110 & 76,503 & 290,861 & 256.203 & 295.974 \\
1 & Perturbed & 116 & 79,911 & 304,773 & 268.939 & 309.799 \\
2 & Loop only & 110 & 76,144 & 290,382 & 257.026 & 296.631 \\
2 & Perturbed & 116 & 79,552 & 304,294 & 268.380 & 308.671 \\
3 & Loop only & 110 & 76,483 & 288,747 & 257.521 & 296.758 \\
3 & Perturbed & 116 & 79,891 & 302,659 & 267.643 & 307.455 \\
\bottomrule
\end{tabular}
\end{table*}

For a target call with $n>1$ output tokens, effective decoding time per output token is $1000(t_{\mathrm{last\ chunk}}-t_{\mathrm{first\ chunk}})/(n-1)$ in milliseconds. The endpoints are client-observed streaming events; the interval includes scheduling and streaming effects. Figure~\ref{fig:eval-cotenancy}(b) averages the paired request differences within each input and block, then averages the three block means for each bar. Its dots retain each block mean. Table~\ref{tab:matched-decode-inputs} gives all six groups; input IDs 1, 2, and 3 follow the binding order above.

\begin{table}[!t]
\centering\scriptsize
\setlength{\tabcolsep}{3pt}
\caption{Paired increases in effective decoding time per output token (ms). Each cell averages three repeated targets of one frozen input within one block.}
\label{tab:matched-decode-inputs}
\begin{tabular}{@{}llrrr@{}}
\toprule
Call & Input & Block 1 & Block 2 & Block 3 \\
\midrule
1 & 1 & 2.654 & 2.405 & 2.011 \\
1 & 2 & 3.068 & 2.717 & 2.704 \\
1 & 3 & 3.075 & 2.711 & 2.651 \\
2 & 1 & 1.868 & 1.590 & 1.503 \\
2 & 2 & 2.094 & 1.872 & 1.552 \\
2 & 3 & 2.293 & 2.376 & 1.945 \\
\bottomrule
\end{tabular}
\end{table}

\begin{figure}[!t]
\centering
\includegraphics[width=\columnwidth]{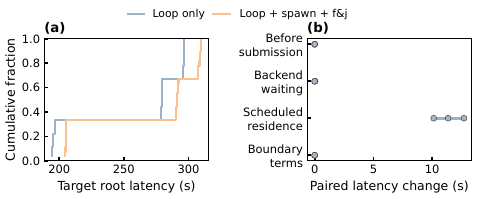}
\vspace{-20pt}
\caption{\textbf{Near-matched fixed-target completion paths.} (a) Target-root latency distributions. (b) Paired changes in block-mean completion-path terms. Each point represents one block; scheduled residence includes pauses after first scheduling.}
\label{fig:app-s65-completion}
\end{figure}

Figure~\ref{fig:app-s65-completion} locates the additional target time in the completion-path ledger. Scheduled residence rises by 12.705, 11.356, and 10.126 seconds across blocks, accounting for almost all target-latency changes. This interval begins at first scheduling and includes subsequent pauses. All six windows have zero KV preemptions and a processed/source prefill factor of 1.0. The separate unmatched-demand sensitivity, including its original background-work and latency ratios, appears in Appendix~\ref{app:interference} and Figure~\ref{fig:app-target-unmatched}.

\subsection{Separate and Mixed Cohorts}
The stage study includes all 26 stage groups and three paired blocks. We reconstruct root latency from its last-finishing operation, recursively choosing each latest-finishing predecessor and recording boundary and scheduling gaps. The complete audit reconciles 1,600 roots and 16,069 operations with sub-microsecond residuals: 1,536 MAS roots form the separate/mixed comparison, and 64 roots belong to an additional ordinary-serving control window. All 219 sequence roots in the same-user comparison have six operations on their observed completion path; their sixteen logical operations overlap across branches.

\begin{figure}[!t]
\centering
\includegraphics[width=\columnwidth]{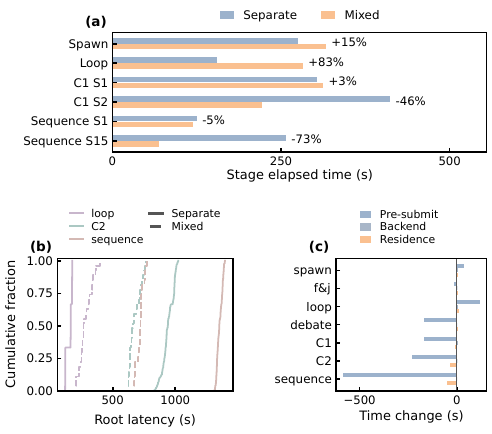}
\vspace{-20pt}
\caption{\textbf{Stage-to-task changes across cohorts.} The original seven-structure Separate/Mixed characterization, Qwen3.5-9B/RTX PRO 5000. (a) Selected stage times and paired percentage changes. (b) Root latency. (c) Mixed-minus-separate path changes; positive means longer latency. Separate runs use 64 users per class; mixed runs use 64 total (ten Spawn and nine per remaining class).}
\label{fig:app-separate-mixed}
\label{fig:eval-stages}
\label{fig:eval-waiting}
\end{figure}

Loop slows to $1.83\times$ separate mean root latency, while Sequence falls to $0.53\times$ (three-block median ratios, balancing means across bindings). Ratios of mean path changes to mean root changes attribute 91.5\% of Loop's penalty and 92.3\% of Sequence's improvement to readiness-to-submission time (Figure~\ref{fig:app-separate-mixed}(c)). This cohort comparison changes both co-execution and per-class load; Figure~\ref{fig:eval-temporal-mixing} instead holds the entire realized workload and global arrival slots fixed.

\begin{figure}[!t]
\centering
\includegraphics[width=\columnwidth]{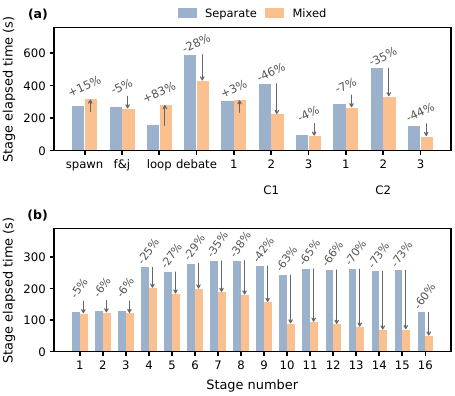}
\vspace{-20pt}
\caption{\textbf{Complete stage response to mixed composition.} Qwen3.5-9B/RTX PRO 5000. Separate runs use 64 users per class; mixed runs use 64 total. All 26 stage groups: bars are medians of three binding-balanced block means; labels are medians of paired block ratios. Short arrows use a minimum display length. C1/C2 follow Figure~\ref{fig:eval-execution}; (b) is the research--review--writing sequence. Figure~\ref{fig:app-separate-mixed}(a) selects six of these groups.}
\label{fig:app-all-stages}
\end{figure}

\begin{figure}[!t]
\centering
\includegraphics[width=\columnwidth]{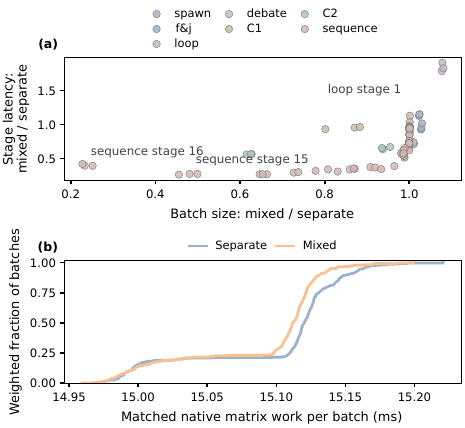}
\vspace{-20pt}
\caption{GPU mixed attribution, Qwen3.5-9B/RTX PRO 5000. (a) all 78 stage/block pairs linking co-batch exposure to stage latency. (b) balanced conditional native matrix-work distributions for early loop-containing batches, matched on the same physical GPU and scheduler conditions.}
\label{fig:app-conditioned}
\end{figure}

Table~\ref{tab:root-accounting} gives selected binding-stratified mean accounts. Figure~\ref{fig:app-conditioned} combines all 78 stage/block exposure pairs with a conditional native comparison. The latter matches physical GPU, phase, request count, scheduled tokens, matrix launch count, and mean-context bins, then balances each common cell by its smaller arm count. The early loop comparison includes 294 separate and 300 mixed batches; its matched matrix-work ratio is 0.999 at a 256-token bin width. Widths 128 and 512 preserve the result. Late loop matching covers about 41\% of mixed batches, and late f\&j has zero common cells. Mapped matrix work belongs to the shared scheduled batch.

\begin{table}[!t]
\centering\scriptsize
\setlength{\tabcolsep}{3pt}
\caption{Mean root-latency changes (mixed minus separate), seconds. Columns follow observed completion paths. Backend and boundary terms complete the exact ledger.}
\label{tab:root-accounting}
\begin{tabular}{@{}lrrr@{}}
\toprule
Workflow & Root & Pre-submit & Residence \\
\midrule
Loop & +129.48 & +118.49 & +11.21 \\
Sequence & -635.39 & -586.36 & -49.05 \\
Spawn--f\&j--loop & -178.56 & -167.44 & -10.93 \\
f\&j--debate--loop & -264.74 & -231.78 & -33.03 \\
\bottomrule
\end{tabular}
\end{table}

\begin{table}[!t]
\centering\scriptsize
\setlength{\tabcolsep}{2pt}
\caption{Root-count ledger across the three blocks. Each separate window contains 64 roots of one class. Each of three mixed windows contains 10 Spawn roots and 9 of each other class. All recorded roots complete. The ordinary control is an additional window in the full audit.}
\label{tab:root-run-ledger}
\begin{tabular}{lrrrrrr}
\toprule
Class & \shortstack{Separate\\runs} & \shortstack{Roots/\\run} & \shortstack{Separate\\total} & \shortstack{Mixed/\\block} & \shortstack{Mixed\\total} & \shortstack{Audit\\total} \\
\midrule
Spawn & 3 & 64 & 192 & 10 & 30 & 222 \\
Fork--Join & 3 & 64 & 192 & 9 & 27 & 219 \\
Loop & 3 & 64 & 192 & 9 & 27 & 219 \\
Debate & 3 & 64 & 192 & 9 & 27 & 219 \\
C1 & 3 & 64 & 192 & 9 & 27 & 219 \\
C2 & 3 & 64 & 192 & 9 & 27 & 219 \\
Sequence & 3 & 64 & 192 & 9 & 27 & 219 \\
MAS total & 21 & -- & 1344 & 64 & 192 & 1536 \\
Ordinary control & 1 & 64 & 64 & -- & -- & 64 \\
\bottomrule
\end{tabular}
\end{table}

\noindent\textbf{Statistical definitions.}
For each arm, workflow, and block, we average latency within each common task binding and then average these binding means with equal weight. Stage bars are the medians of these block means; stage labels summarize the median of the three paired mixed/separate ratios. The reported root ratios follow the same mean-then-ratio procedure, summarized by the median across blocks. Root distribution curves retain individual observed latencies. The time ledger instead averages paired differences of binding-balanced means across blocks. Thus, the 91.5\% and 92.3\% figures divide mean pre-submission changes by mean root-latency changes under the same block and binding aggregation.

The ledger retains IDs, bindings, counts, and source hashes for all 25 windows: 21 separate, 3 mixed, and 1 ordinary control. Cohorts share task bindings with different per-class populations. Pre-submission spans readiness to submission; dependency-release and boundary gaps remain separate terms.

\begin{table}[!t]
\centering\footnotesize
\setlength{\tabcolsep}{1.75pt}
\caption{All 26 stage groups underlying Figure~\ref{fig:app-separate-mixed}. S/M are separate/mixed binding-balanced stage means in seconds; columns retain every paired block.}
\label{tab:all-stage-blocks}
\begin{tabular}{llrrrrrr}
\toprule
Workflow & Stage & S0 & M0 & S1 & M1 & S2 & M2 \\
\midrule
Spawn & S1 & 275.73 & 311.09 & 275.35 & 316.60 & 277.97 & 319.44 \\
Fork--Join & S1 & 272.41 & 252.68 & 270.12 & 256.39 & 269.67 & 274.15 \\
Loop & S1 & 154.64 & 276.15 & 153.14 & 292.11 & 154.54 & 282.51 \\
Debate & S1 & 593.07 & 428.92 & 591.09 & 419.20 & 590.47 & 437.42 \\
C1 & S1 & 302.02 & 341.09 & 302.86 & 312.72 & 305.40 & 298.56 \\
C1 & S2 & 412.10 & 211.89 & 411.93 & 221.74 & 409.25 & 235.69 \\
C1 & S3 & 93.00 & 86.72 & 92.30 & 88.16 & 92.97 & 89.57 \\
C2 & S1 & 285.00 & 252.92 & 283.67 & 267.58 & 283.82 & 263.20 \\
C2 & S2 & 508.42 & 340.65 & 507.17 & 323.06 & 505.57 & 330.94 \\
C2 & S3 & 151.40 & 85.54 & 152.04 & 85.06 & 151.56 & 85.47 \\
Sequence & S1 & 125.27 & 106.88 & 125.17 & 120.59 & 126.43 & 119.85 \\
Sequence & S2 & 129.30 & 110.33 & 129.64 & 124.14 & 130.36 & 122.64 \\
Sequence & S3 & 127.99 & 108.20 & 127.93 & 122.13 & 128.30 & 120.98 \\
Sequence & S4 & 266.66 & 195.74 & 267.69 & 202.69 & 266.21 & 200.72 \\
Sequence & S5 & 254.09 & 176.22 & 250.67 & 182.42 & 252.03 & 182.90 \\
Sequence & S6 & 282.10 & 191.09 & 278.94 & 198.33 & 279.06 & 202.67 \\
Sequence & S7 & 287.24 & 187.80 & 288.65 & 177.17 & 286.21 & 187.41 \\
Sequence & S8 & 287.14 & 184.48 & 290.69 & 168.80 & 288.07 & 179.41 \\
Sequence & S9 & 270.94 & 167.86 & 274.16 & 152.47 & 271.07 & 155.97 \\
Sequence & S10 & 244.22 & 94.88 & 239.23 & 87.85 & 242.15 & 83.28 \\
Sequence & S11 & 261.67 & 98.06 & 265.25 & 92.65 & 262.25 & 92.79 \\
Sequence & S12 & 258.46 & 87.57 & 254.12 & 86.23 & 260.97 & 80.53 \\
Sequence & S13 & 261.42 & 83.21 & 261.28 & 76.40 & 258.01 & 77.42 \\
Sequence & S14 & 255.05 & 68.81 & 251.47 & 67.64 & 253.85 & 68.19 \\
Sequence & S15 & 257.51 & 70.31 & 254.49 & 69.29 & 260.35 & 68.16 \\
Sequence & S16 & 124.23 & 49.11 & 124.18 & 48.85 & 117.15 & 49.16 \\
\bottomrule
\end{tabular}
\end{table}

\subsection{Unmatched Background Demand} 
\begin{figure}[!t]
\centering
\includegraphics[width=\columnwidth]{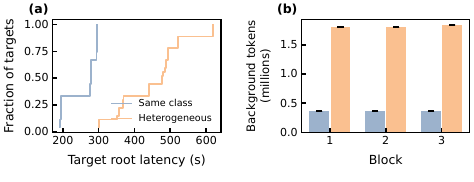}
\vspace{-20pt}
\caption{\textbf{Unmatched background demand on fixed targets.} (a) All 27 target-root latencies per arm. (b) Background input/output work by block. The heterogeneous arm supplies substantially more work.}
\label{fig:app-target-unmatched}
\end{figure}

Figure~\ref{fig:app-target-unmatched} retains the earlier Qwen3-8B/910C comparison at request limit 128. Each arm has nine identical tagged Loop targets and 55 background roots. Target IDs, bindings, arrivals, resolved payloads, output lengths, and dependencies match across the same-class and heterogeneous backgrounds. All 27 targets per arm complete across three blocks. Heterogeneous backgrounds supply $4.92\times$ as many input/output tokens and raise mean target latency to $1.77\times$ the same-class value (medians of paired-block ratios). Table~\ref{tab:target-background} gives this unmatched-demand ledger, distinct from the approximately matched perturbation in Appendix~\ref{app:matched-background}.

\begin{table}[!t]
\centering\scriptsize
\setlength{\tabcolsep}{3pt}
\caption{Unmatched-demand fixed-target run ledger. Root time is the mean over nine tagged targets; background work sums source input and output tokens across 55 roots.}
\label{tab:target-background}
\begin{tabular}{@{}llrr@{}}
\toprule
Block & Background & Target mean (s) & Background tokens \\
\midrule
1 & Same class & 254.946 & 367,364 \\
1 & Heterogeneous & 451.322 & 1,808,641 \\
2 & Same class & 256.105 & 366,526 \\
2 & Heterogeneous & 411.475 & 1,798,945 \\
3 & Same class & 257.084 & 365,230 \\
3 & Heterogeneous & 491.703 & 1,838,309 \\
\bottomrule
\end{tabular}
\end{table}

\subsection{Earlier Fixed-Population Temporal Mixing}
\label{app:temporal-mixing}
This comparison reuses nine tagged Loop targets and the heterogeneous 55-root background from the near-matched study, keeping the entire population fixed across each pair. The background contains 53 Loop roots, one Spawn, and one Fork--Join. Canonical requests, output lengths, task bindings, operation IDs, typed dependencies, tools, class counts, and target arrivals remain identical. Three blocks retain 134 calls each, with input/output totals of 92,409/352,227, 92,050/351,748, and 92,389/350,113 tokens, respectively, equal between paired arms.

A pre-execution manifest maps lexicographic classes and stable root IDs to the same 55 slots by round-robin interleaving (A) or class concatenation (B). Targets retain their arrivals; Spawn moves from 1.26 to 0.22 seconds and Fork--Join retains 0.18 seconds. No ordering is selected using outcomes.

Qwen3-8B/910C pairs use BF16, TP1, request limit 128, 31.99\,GiB KV, and prefix caching off, with A/B, B/A, A/B order and empty paired boundaries. All six cohorts complete 64 roots each; all 804 requests satisfy their contracts, with zero KV preemption. Task/path definitions are unchanged.

Table~\ref{tab:temporal-mixing} retains all blocks: ratios of nine-target arithmetic means yield median $-0.246\%$ and range $[-0.796,+0.348]\%$, mainly in scheduled residence. The ledger retains all 27 pairs. Figure~\ref{fig:eval-temporal-mixing} uses the balanced population in Appendix~\ref{app:balanced-temporal-mixing}.

\begin{table}[!t]
\centering\scriptsize
\setlength{\tabcolsep}{2.2pt}
\caption{Fixed-population temporal mixing. A/B is interleaved/clustered mean target latency; deltas are A minus B. Pre: ready to submission; Wait: submission to first scheduling; Res: first scheduling to completion, including pauses.}
\label{tab:temporal-mixing}
\begin{tabular}{@{}rrrrrr@{}}
\toprule
Block & A/B & $\Delta$Root (s) & $\Delta$Pre (ms) & $\Delta$Wait (ms) & $\Delta$Res (s) \\
\midrule
1 & 0.9975 & -0.657 & -0.005 & -6.73 & -0.651 \\
2 & 1.0035 & +0.924 & -0.002 & -5.14 & +0.930 \\
3 & 0.9920 & -2.124 & +0.042 & +9.33 & -2.134 \\
\bottomrule
\end{tabular}
\end{table}

\end{document}